# The Low-mass Stellar Initial Mass Function in Nearby Ultrafaint Dwarf Galaxies

Carrie Filion[1], Rosemary F. G. Wyse[1], Hannah Richstein[2], Nitya Kallivayalil[2], Roeland P. van der Marel[1,3], and Elena Sacchi[4]

[1] Department of Physics & Astronomy, The Johns Hopkins University, Baltimore, MD 21218, USA; cfilion@jhu.edu  
[2] Department of Astronomy, The University of Virginia, 530 McCormick Road, Charlottesville, VA 22904, USA  
[3] Space Telescope Science Institute, 3700 San Martin Drive, Baltimore, MD 21218, USA  
[4] Leibniz-Institut für Astrophysik Potsdam, An der Sternwarte 16, 14482 Potsdam, Germany  
*Received 2024 February 23; revised 2024 April 16; accepted 2024 April 16; published 2024 May 30*

## Abstract

The stellar initial mass function (IMF) describes the distribution of stellar masses that form in a given star formation event. The long main-sequence lifetimes of low-mass stars mean that the IMF in this regime (below $\sim 1\,M_\odot$) can be investigated through star counts. Ultrafaint dwarf galaxies are low-luminosity systems with ancient, metal-poor stellar populations. We investigate the low-mass IMF in four such systems (Reticulum II, Ursa Major II, Triangulum II, and Segue 1), using Hubble Space Telescope imaging data that reaches to $\lesssim 0.2\,M_\odot$ in each galaxy. The analysis techniques that we adopt depend on the number of low-mass stars in each sample. We use Kolmogorov–Smirnov tests for all four galaxies to determine whether their observed apparent magnitude distributions can reject a given combination of IMF parameters and binary fraction for the underlying population. We forward model 1000 synthetic populations for each combination of parameters, and reject those parameters only if each of the 1000 realizations reject the null hypothesis. We find that all four galaxies reject a variety of IMFs, and the IMFs that they cannot reject include those that are identical, or similar, to that of the stellar populations of the Milky Way. We determine the best-fit parameter values for the IMF in Reticulum II and Ursa Major II and find that the IMF in Reticulum II is generally consistent with that of the Milky Way, while the IMF in Ursa Major II is more bottom heavy. The interpretation of the results for Ursa Major II is complicated by possible contamination from two known background galaxy clusters.

*Unified Astronomy Thesaurus concepts:* Low mass stars (2050); Initial mass function (796); Dwarf spheroidal galaxies (420); Local Group (929)

## 1. Introduction

The stellar initial mass function (IMF) describes the mass distribution of single stars formed in any given star formation event. The IMF plays a critical role in the cosmic baryon cycle and has far-reaching implications for many subfields of astrophysics. The shape of the IMF— both within the Milky Way and in external galaxies—has been the subject of debate for decades (as reviewed by, e.g., Bastian et al. 2010), and the physical mechanisms underlying the IMF remain ill understood. The different stellar components of the Milky Way are generally consistent with having formed with the same IMF, independent of age or metallicity, for example. Whether or not this invariance extends to external galaxies continues to be debated (e.g., van Dokkum & Conroy 2021); the IMF in ultrafaint dwarf (UFD) galaxies[5] is of particular interest, as these low-luminosity, dark-matter dominated galaxies made of low-metallicity, ancient stars probe an extreme of both star and galaxy formation. The low-mass IMF is accessible through star counts, and in this analysis, we build upon our previous work (Filion et al. 2022, hereafter Paper I), in which we determined the low-mass IMF in Boötes I (Boo I)—finding consistency with that of the Milky Way—with an analysis of four more UFD galaxies, namely, Reticulum II, Ursa Major II, Triangulum II, and Segue 1. Of these, only Reticulum II has a previously published estimate of the low-mass IMF (Safarzadeh et al. 2022), derived from shallower Hubble Space Telescope (HST) imaging data than those analyzed in this paper, and which was obtained with the primary science goal of determining the wide-binary population. We discuss those results, together with those obtained by other authors for different UFD galaxies, in Section 5.

We first describe our sample of UFD galaxies and their associated data in Section 2.1 below, then give details of the creation of the photometric catalogs of likely members (Section 2.2). We then introduce the methods we used to constrain the low-mass IMF in each galaxy in Section 3, and present the results of these analyses in Section 4. We discuss our findings in the context of previous studies in Section 5 and present our conclusions in Section 6.

## 2. Observational Data

### 2.1. Selection of Ultrafaint Galaxies

We selected the sample of UFD galaxies for this analysis based on the requirement that they have suitable deep optical HST imaging data available from the Barbara A. Mikulski Archive for Space Telescopes (MAST), plus a sufficient number of stars with good photometry to avoid issues of small number statistics. The level of uniformity of the data quality across the sample of UFD galaxies is an important factor in the robustness of intercomparisons, so we focused on observing programs that targeted several UFD galaxies.

---

[5] Typically defined to be less luminous than $\sim 10^5\,L_\odot$ (e.g., Simon 2019).

The stellar mass range over which the low-mass IMF can be constrained is critical to the interpretation of the results. Our main science goal is a comparison with the IMF that has been derived for the stellar populations in the Milky Way; that IMF is typically described by either a lognormal function with characteristic mass around $0.2\,\mathcal{M}_\odot$ (e.g., Chabrier 2005) or by a broken power law, with a flatter slope for masses below around $0.5\,\mathcal{M}_\odot$ (e.g., Kroupa 2001). The lower-mass limit of an intended comparison data set should be at, or below, the mass at which the Milky Way IMF turns over or flattens (as emphasized by El-Badry et al. 2017), which in turn sets the required faint limit of the photometric data for each UFD galaxy to be the magnitude corresponding to $\sim 0.2\,\mathcal{M}_\odot$. The effective upper-mass limit of a star-count based analysis of the IMF of stars still on the main sequence in UFD galaxies is set by the oldest age of the member stars, and is $\lesssim 1\,\mathcal{M}_\odot$.

We first limited the search to UFD galaxies closer than 50 kpc, the expected distance limit within which it is feasible that deep HST imaging could probe masses $\sim 0.2\,\mathcal{M}_\odot$ (see Figure 7 of El-Badry et al. 2017, for the example of HST Advanced Camera for Surveys, hereafter ACS, F814W data), although most UFD galaxies discovered thus far satisfy this limit. We also required that the data provide a signal-to-noise ratio equal to at least 5 at the apparent magnitude corresponding to a $\sim 0.2\,\mathcal{M}_\odot$ star, based on expectations from the appropriate HST exposure time calculator.

We identified the HST ACS Treasury Program GO-14734 (PI: N. Kallivayalil) as the most promising source and selected Reticulum II (Ret II), Ursa Major II (UMa II), Triangulum II (Tri II), and Segue 1 (Seg 1) as our sample of UFD galaxies. Each of these has sufficiently deep imaging and a large enough stellar population (quantified in Section 3.1 below) for our analysis of the low-mass IMF. Ret II is plausibly a satellite of the Magellanic Clouds (e.g., Patel et al. 2020), while the rest of the sample are likely satellites of the Milky Way, allowing an investigation into the possible role of local environment/host galaxy. The chemical abundances of Ret II member stars are also known to be unusual, in that a large fraction of Ret II stars show a significant enhancement in $r$-process elements (Ji et al. 2016).

The imaging data for each galaxy were taken in the F814W and F606W filters, with the ACS Wide Field Channel, which has a field of view of $202'' \times 202''$. The total exposure time in each filter, for each field center (pointing), was $\sim 4600$ s, split over two orbits per filter. UMa II, Tri II, and Seg 1 were imaged using two separate pointings, while the relatively brighter Ret II was imaged using just one pointing. The individual exposures in each pointing (four per filter) were dithered with a four-point dither pattern.

## 2.2. Creation of Photometric Catalogs

The individual exposures for each field were aligned and then combined using the HST software package `Drizzle-Pac`.[6] The drizzled output images (referred to below as "drc images") were masked using the `segmentation` routine within the astropy package `photutils` (Bradley et al. 2023), in order to minimize possible adverse effects on the photometry from the detector chip gaps and extended saturated sources. Here, we give only a brief overview of the photometry pipeline, with a more comprehensive description being provided in H. Richstein et al. (2024, in preparation). The photometric catalogs for each of the galaxies in the GO-14734 Treasury Program—including those discussed here—will be provided electronically on MAST upon publication of the comprehensive description (the catalogs will be accessible via doi:10.17909/b5gn-6e22, and the imaging data used in this analysis are available via doi:10.17909/mwdx-da66). We performed point-spread function (PSF)-fitting photometry on the masked F606W and F814W drc images, using DAOPHOT-II and ALLSTAR (Stetson 1987, 1992). The resulting source lists were input into DAOMATCH and DAOMASTER, to create a preliminary PSF source catalog. The magnitudes were corrected to account for the finite PSF aperture, calibrated to the VegaMag system, and adjusted for the exposure time. The 50% completeness limit in each filter was determined through artificial star tests, and the resulting values for each UFD galaxy are given in Table 1, indicated by the subscript "max."

We removed non-star-like objects via cuts on the quality-of-fit statistics determined for each source during the PSF-fitting process. For each target, we fit exponential functions to three quality-of-fit parameters, then used the functional fit with a $3\sigma$ spread to divide between acceptable and poor values of each parameter. Each photometric source was assigned a quality flag with a value between 0 and 1, based on our assessment of the three quality-of-fit parameters in each of the two filters. For more details on this flag, we refer the reader to Richstein et al. (2024; see also H. Richstein et al. 2024, in preparation). We required acceptable sources to have flag values $\geqslant 0.7$, and those sources that passed these image-quality tests are referred to below as "star-like sources," albeit that they will inevitably contain faint, barely resolved extragalactic contaminants. Indeed, the resulting catalogs of star-like sources for each UFD galaxy pointing are a mix of actual members of that UFD galaxy—both single stars and unresolved binary systems—together with Milky Way stellar contaminants and the above-mentioned faint background galaxies. In principle, we could include such contaminants in our forward modeling of the photometric data, but we lack the required detailed models of the structure of the stellar populations of the Milky Way (existing models lack substructure) and of the high-redshift galaxy population. We also do not know what fraction of these sources will make it through the various criteria applied by our photometric pipeline. However, significant parts of each of these two classes of contaminant sources have different distributions of color and apparent magnitude than do the UFD galaxy member sources, and thus, a large fraction of each can be removed from the photometric catalogs through color–magnitude cuts, as described next.

### 2.2.1. Identification of Candidate Member Sources

Existing photometric and spectroscopic data for each of the UFD galaxies in our sample are consistent with stellar populations that are uniformly old and metal poor, albeit with a significant internal metallicity dispersion about the well-defined mean (see Table 1). This relative simplicity allows us to select a fiducial theoretical isochrone for each UFD galaxy that describes the expected (mean) locus of the member stars in the color-apparent magnitude plane. We chose to use the isochrones from the Dartmouth Stellar Evolution Database (Dotter et al. 2008) for this purpose, as they provide "Equal Evolutionary Points" (see, e.g., Dotter 2016; note that the meaning of the Equal Evolutionary Points in their isochrone

---

[6] https://www.stsci.edu/scientific-community/software/drizzlepac

**Table 1**
Galaxy Sample

| Galaxy | ℓ (deg) | b (deg) | Distance (kpc) | $A_V$ | $M_V$ | F814W$_{max}$ | F814W$_{min}$ | F606W$_{max}$ | Reference |
|---|---|---|---|---|---|---|---|---|---|
| Reticulum II | 266 | −50 | 31.6 | 0.05 | −4.0 | 27.06 | 21.00 | 28.02 | (1, 2) |
| Ursa Major II | 153 | +37 | 34.7 | 0.25 | −4.4 | 27.05 | 21.50 | 27.91 | (1, 3) |
| Triangulum II | 141 | −24 | 28.4 | 0.22 | −1.6 | 26.88 | 20.90 | 27.70 | (1, 2) |
| Segue 1 | 221 | +50 | 22.9 | 0.08 | −1.3 | 26.96 | 20.50 | 27.80 | (1, 3, 4) |

**Note.** For each galaxy, from left to right, we list the name, galactic coordinates (ℓ and b) in degrees, the distance in kiloparsecs, the extinction in the V band, the absolute magnitude of the galaxy, the faint and bright magnitude limits in the F814W filter (F814W$_{max}$ and F814W$_{min}$, respectively), and the faint limit in the F606W filter (F606W$_{max}$). References: (1) absolute magnitudes and distances taken from the compilation in Simon (2019), see references therein; note that an IMF has been assumed to estimate the total luminosity; (2) Sacchi et al. (2021) and references therein; (3) $A_V$ value from Schlafly & Finkbeiner (2011), provided by the NASA IPAC IRSA Galactic Dust and Extinction web interface; (4) distance from the compilation in McConnachie (2012).

library is the same as in the library adopted here), making it easy to interpolate within them, in addition to giving a good overall match to the photometric data. We assumed an age of 13 Gyr for all of the UFD galaxies and selected the metallicity of each solar-scaled isochrone based on the spectroscopic data, as summarized in Table 1, albeit that for most we adopted the lowest metallicity available for these isochrones, which is −2.5 dex. We implemented color cuts around the fiducial isochrone for each galaxy as a means of distinguishing likely member star-like sources from nonmembers.

The fiducial isochrones were first transformed to apparent magnitudes by adopting the distances given in Table 1, and then shifted to account for the effects of dust extinction in each band. We used the Spanish Virtual Observatory (SVO) filter-profile service[7] to convert the $A_V$ values in Table 1 to the F814W and F606W passbands. We confirmed that the 50% completeness limit in the F814W filter for each galaxy (given as F814W$_{max}$ in Table 1) corresponds to a mass below $0.2\mathcal{M}_\odot$ on the fiducial isochrone.

Milky Way white dwarf stellar remnants and semi-resolved high-redshift background galaxies will generally be bluer than the member sources of each galaxy, while stars in the Milky Way occupy a broad range of color and apparent magnitude, including the locus of the member stars of each galaxy. We implemented color cuts as a function of apparent magnitude to remove most of these potential contaminants, using the fiducial isochrone as a guide. The width in color about the fiducial isochrone was determined by the photometric errors at the apparent magnitude corresponding to the 50% completeness limit. To be specific, we first computed the mean error in color for the star-like sources within ±0.05 mag of the F814W 50% completeness limit, by adding the F606W and F814W magnitude errors in quadrature and then taking the mean. The resulting mean error is approximately 0.1 mag. We then defined nonmember star-like sources to be those with colors that are bluer (redder) than the galaxy's fiducial isochrone by more than 3 times (4 times) this mean error in color, where the less stringent red cut allows for the presence of unresolved binary star systems. We further retained as candidate members only those remaining star-like sources that were brighter than the F606W$_{max}$ magnitude and within the F814W magnitudes given in Table 1, where the bright limit was set slightly fainter than the main-sequence turnoff to reduce the sensitivity to age uncertainties, and the faint limit is the 50% completeness limit in that band.

We also checked for possible contamination from moderate redshift galaxies by querying the SIMBAD astronomical database (Wenger et al. 2000) for known galaxy clusters located within a 10′ radius of the central coordinates of each of the UFD galaxies. Despite the fact that the location on the sky of Ret II lies within the coordinates of the mapped extent of the Horologium-Reticulum Supercluster (Fleenor et al. 2006) and within 10′ of the coordinates given by Kalinkov & Kuneva (1995) for the Supercluster labeled 225, Ret II is not within 10′ of any of the constituent galaxy clusters. Similarly, no previously identified galaxy cluster lies within this projected distance of either Seg 1 or Tri II. There are, however, two galaxy clusters this close to UMa II, both at redshift $z \sim 0.5$. The locations of the three brightest galaxies in each cluster, reported in Szabo et al. (2011), are indicated in the left-hand panel of Figure 1, together with the detected sources in the UMa II field, while their extinction-corrected colors and magnitudes are compared to the fiducial isochrone of UMa II in the right-hand panel. It is clear that the luminous red galaxies, which tend to be centrally concentrated within clusters, are not important contaminants, but the more numerous fainter (and bluer) galaxies in the clusters could possibly lie within the color cuts we implemented. The level of possible contamination is difficult to quantify, and we will revisit this below, when we analyze the data for UMa II.

It is difficult—if not impossible—to remove Milky Way stars that occupy the same locus in color–magnitude space as the true members of a given UFD galaxy without additional information, such as proper motions. However, star-count models of the Milky Way can be used to estimate the level of such contamination, and happily, it is expected to be small. For example, the TRILEGAL model (Girardi et al. 2005) predicts that in the lines of sight of our sample galaxies there are ∼10 Milky Way stars per pointing that lie within the color and magnitude ranges defined for member stars, corresponding to a residual contamination level of lower than 5% for each galaxy. We therefore refer below to all the star-like sources that pass the magnitude and color cuts as "member sources." Their locations in the color (apparent) magnitude plane are shown in Figure 2, where the likely member sources are shown with filled circles, and nonmembers are shown with gray crosses.

## 3. Constraining the Low-mass Initial Mass Function

The aim of this analysis is to constrain the single-star IMF, with each star in the underlying (unresolved) binary population explicitly incorporated into the modeling approach, rather than the system IMF, which counts binaries as a single source. There are two widely adopted functional forms of the low-mass, single-star stellar IMF that we will use in our analysis, namely, (i) a lognormal function and (ii) a two-segment (broken) power law. The first of these is exemplified by the Chabrier IMF, obtained in Chabrier (2003). As updated in Chabrier (2005), the number of low-mass stars per logarithmic interval in mass is fit with a characteristic mass, $M_{cc}$, equal to $0.2\mathcal{M}_\odot$, and a width parameter, $\sigma_C$, equal to 0.55. The broken power-law form is exemplified by the Kroupa IMF (Kroupa 2001), where the break occurs at $M_{BK} = 0.5\mathcal{M}_\odot$, and adopting the definition of IMF in terms of number of stars in linear mass intervals, the slopes below and above this break mass are, respectively, $\alpha_{1K} = -1.3$ and $\alpha_{2K} = -2.3$. These functional forms exhibit a downturn at low mass and supersede the single power-law IMF found in the seminal analysis of Salpeter (1955, power-law slope of $\alpha_S = -2.35$). The Salpeter IMF was originally constrained over the relatively high-mass range of $0.4\,\mathcal{M}_\odot$–$10\,\mathcal{M}_\odot$ and assuming it holds down to lower masses overpredicts the number of low-mass stars compared to modern data. We will however use this slope as a comparison point in the analysis below. Throughout the text, we use subscripts (C, K, and S for Chabrier, Kroupa, and Salpeter, respectively) on parameters to indicate these "canonical" values.

---
[7] Available at http://svo2.cab.inta-csic.es/theory/fps/.

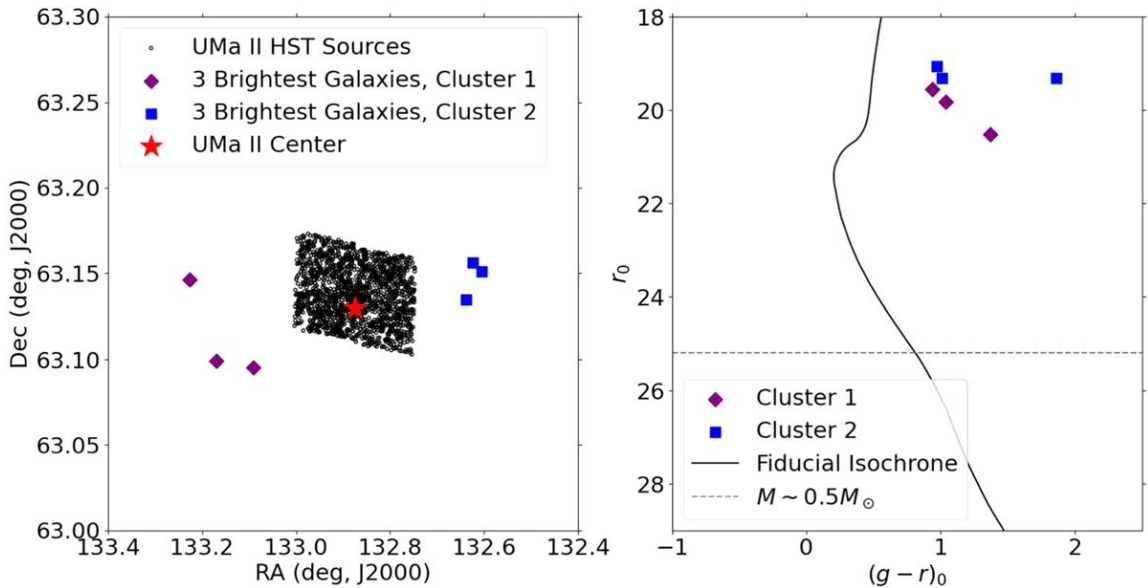

**Figure 1.** Left: the location on the sky of the three brightest galaxies in the two clusters, taken from Szabo et al. (2011), together with the detected sources in the UMa II field. Right: the extinction-corrected colors and apparent magnitudes of the three brightest galaxies in each cluster, plus the fiducial isochrone for UMa II in the SDSS passbands, with the approximate apparent magnitude of a $0.5\mathcal{M}_\odot$ star at the distance of UMa II indicated by the horizontal dashed line.

The Chabrier and Kroupa IMFs are generally valid above the hydrogen burning limit ($\sim 0.08 M_\odot$), and the IMF of brown dwarfs is usually considered separately. The IMF of such extremely low-mass objects is generally thought to be shallower than the IMF above the hydrogen burning limit (for example, the analysis of the local disk in Kirkpatrick et al. 2024; or a young cluster in Mužić et al. 2017), although there remains uncertainty as to exactly at what mass the slope changes. Further, the very faint nature of brown dwarfs tends to limit analyses to relatively small samples, which in turn limits the precision with which the IMF can be determined, as discussed below.

### 3.1. Minimum Sample Size

Obtaining reliable constraints on the underlying IMF of a given stellar population requires sufficiently deep imaging data and a suitably large number of stars. Data that probe only a limited mass range can be consistent with a variety of IMF parameters, as discussed in El-Badry et al. (2017), although those authors demonstrate that a larger sample size can improve the discriminatory power of the analysis. The combination of a large sample with data sensitive to as wide as possible a mass range is ideal. For example, El-Badry et al. (2017) find that testing whether or not the underlying IMF is consistent with the Chabrier lognormal IMF requires data that have a lower-mass limit at, or below, the characteristic mass of that IMF, $M_{cc} \sim 0.2\,\mathcal{M}_\odot$.

However, even in the regime in which observations reach such very low masses, it is still possible to have too few stars to provide robust constraints on the IMF, which we now demonstrate (see also Weisz et al. 2013; and El-Badry et al. 2017 for further discussion). We first made the (reasonable) assumptions that the low-mass end of the IMF was well sampled during star formation and that mass segregation within the UFD galaxy has not occurred, so that the area studied is representative of the system (see, e.g., Geha et al. 2013; Gennaro et al. 2018a). Next, we generated mock data sets by drawing stellar masses from a Kroupa broken power-law IMF over the mass range between $0.11\mathcal{M}_\odot$ and $1\mathcal{M}_\odot$ (matching the range considered in Section 3.2) to produce samples of sizes ranging from 500 to 10,000 individual stellar masses. For this exercise, we assumed that all stars are single (i.e., no binaries). The statistical uncertainty associated with a given sample size (denoted by $n$) is illustrated graphically in the left-hand panel of Figure 3, which shows the median, 10th, and 90th percentiles of the counts in twenty, linearly spaced mass bins for 1000 realizations of draws from this IMF of size $n$. The area enclosed between the 10th to 90th percentiles increases as $n$ decreases, such that a range of different IMF slopes could be compatible with the mock data, as indicated in the figure. The right-hand panel of Figure 3 shows the 10th and 90th percentiles of the percent difference between a given realization of $n$ masses drawn from the IMF and the 999 additional realizations of this sample size. The impact of small number statistics is evident—for a relatively small sample size, the number of masses in a given mass bin can be quite different from realization to realization, limiting the ability to discriminate between different IMFs, even in this idealized case with only single stars of known mass.

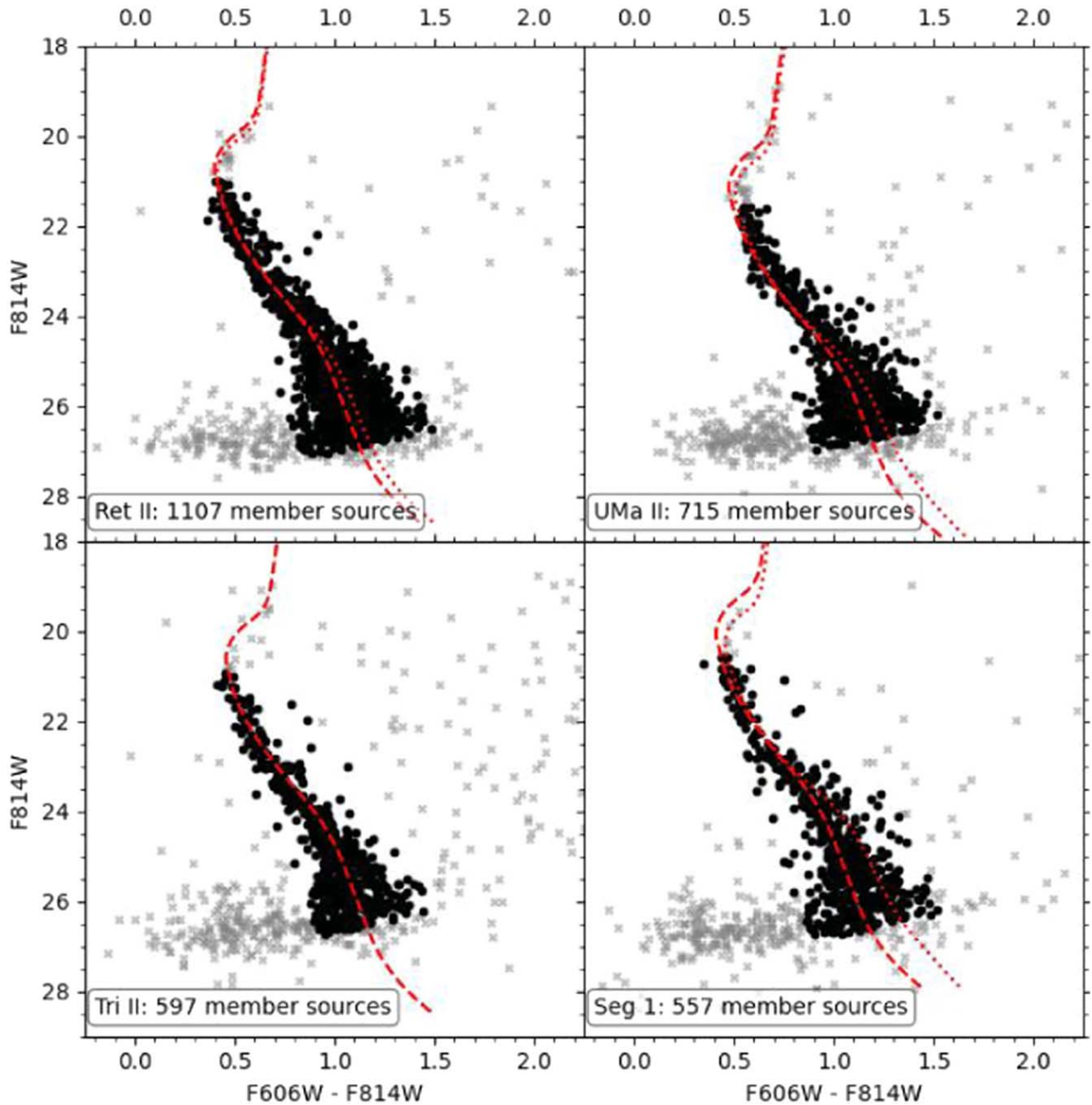

**Figure 2.** Apparent color–magnitude diagrams showing all the star-like sources in the fields of each of the four galaxies analyzed in this study, together with the fiducial isochrone for each (red dashed line), after it has been adjusted for distance and extinction in each band. The isochrones indicated by red dotted lines correspond to the maximum metallicity (see Table 1; note that the fiducial and maximum metallicity for Tri II are the same) and illustrate the effect of the internal metallicity spread. We assume an age of 13 Gyr for each isochrone. The likely member sources—as determined following the procedure described in the text and further restricted to be within the apparent magnitude limits given in Table 1—are indicated by filled black circles while the likely nonmember star-like sources are indicated by gray crosses. The magnitude limits that we imposed at the bright end are slightly fainter than the main-sequence turnoffs; these limits were chosen to reduce the sensitivity of our analyses to uncertainties in the estimated star formation histories and mean stellar ages.

The actual photometric data sets for the UFD galaxies contain between ∼500 and ∼1000 sources, reflecting the area observed and the distance of each galaxy, as well as the intrinsically faint nature of UFD galaxies. It is clear from the results shown in Figure 3 that, for small samples, direct fitting of IMFs to the star counts, even in the idealized limit of error-free, perfect transformation of apparent magnitudes to stellar masses, would not provide a robust solution. We instead adopted two different approaches to investigate the underlying IMF, with the selection between the two being dependent on the sample size: for the larger samples, we were able to provide best-fit values for the IMF parameters through forward modeling and an Approximate Bayesian Computation Markov Chain Monte Carlo (ABC-MCMC) analysis (see Section 3.3), while, for smaller sample sizes, we did not attempt to constrain the best-fit IMF parameter values and instead mapped the range of values that can be rejected by the data, by means of Kolmogorov–Smirnov (KS) tests. We chose an observed sample size ($n_{\mathrm{obs}}$) of 600 star-like member sources as the threshold for the ABC-MCMC approach, guided in part by the fact that ∼60% of masses drawn at random from a Milky Way–like IMF will be main-sequence stars above the mass limit of the observational data, so that an observed sample of 600 represents a parent population numbering ≳1000. Ret II and UMa II both have $n_{\mathrm{obs}} > 600$, and we constrained the IMF of these galaxies using the ABC-MCMC methodology developed in Paper I and summarized below in Section 3.3. The second category of observed UFD galaxies consists of Seg 1 and Tri II, and the process by which we determined the ranges of rejected IMF parameters for these two systems (plus the two in the former category) is described in Section 3.4.

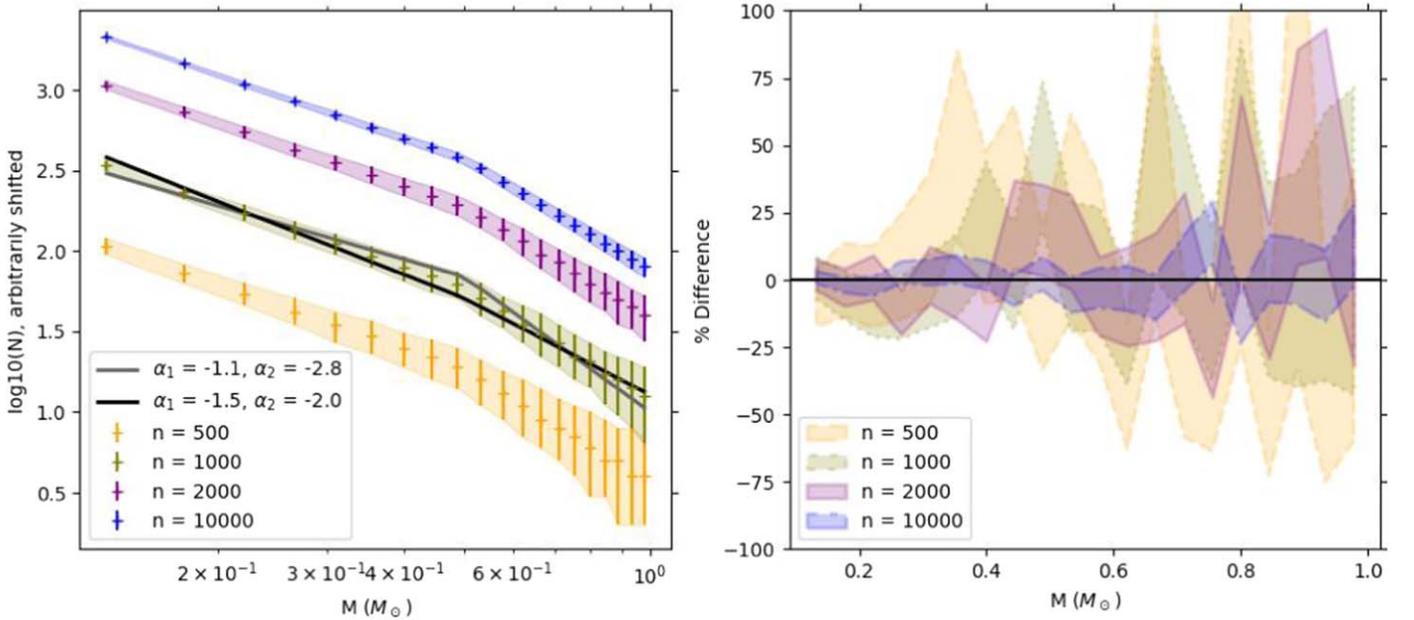

**Figure 3.** Left-hand panel: The distribution of counts in 20 equally spaced (linear) bins in mass, over the range 0.11 $\mathcal{M}_\odot$–1.0 $\mathcal{M}_\odot$, for sample sizes $n = 500$, 1000, 2000, and 10,000 (the normalization is chosen for clarity, with the distribution for increasing $n$ shown from bottom to top, with corresponding colors yellow, green, purple, blue), all drawn 1000 times from the Kroupa power-law IMF. The points indicate the median number in each mass bin, with the error bars and shaded regions corresponding to the 10th and 90th percentiles. Two alternative broken power-law IMFs that lie within the range of the $n = 1000$ yellow shading are indicated by the solid gray and black lines. Right-hand panel: The percentage differences across the 1000 realizations of each sample size, calculated relative to the results for the initial drawing of $n$ masses from the IMF. The area between the 10th and 90th percentile of the percent differences is shaded following the same color-coding as the left-hand panel.

### 3.2. Generation of Synthetic Photometric Data Sets

While we used different approaches to investigate the low-mass IMF of the galaxies with observed sample sizes above and below our required minimum for the determination of the best-fit IMF ($n_{\mathrm{obs,min}} = 600$), both approaches are rooted in forward modeling of synthetic stellar populations. Forward modeling allows us to compare the real, observed data to a synthetic population "observed" in the same parameter space, where the synthetic data were generated under the assumption that the stars formed following a given single-star IMF and with a given binary fraction. This technique requires the output from stellar evolution models, implicitly assuming error-free mass-to-luminosity transformations, together with externally constrained parameters, such as distance and mean stellar age. We utilized unbinned luminosity functions, in the form of the apparent magnitude distributions, in making the comparisons between model and data for all galaxies in our sample, while the use of the Hess diagram (i.e., a binned color–magnitude diagram, see Section 3.3) in the comparison was restricted to the two galaxies with a sufficiently large sample size. The procedure we used to generate the synthetic photometric data and the necessary "observed" quantities was described in detail in Paper I and is outlined below.

The first step is to create a synthetic stellar population for each galaxy, which we modeled under the assumption of a single age—taken to be 13 Gyr—and a range of metallicities. This assumption is reasonable given the typical derived star formation histories of UFD galaxies (e.g., Brown et al. 2014; Sacchi et al. 2021), plus the fact that, at a given (low) metallicity, the isochrones predict that low-mass main-sequence stars of different (ancient) age differ little. The metallicity distribution within each system was based on the available spectroscopic data, as summarized in Table 2 and described in Section 3.2.1 below. We assigned a mass to each "star" by drawing a set of masses from an assumed IMF. As discussed in Paper I, the number of masses that must be drawn in order to reproduce the observational data will always be larger than the number of sources in that data set. The binary fraction, denoted by $f_f$ in the following analysis, was defined as the fraction of all "stars" that are in binary systems.[8] Each "star" was assigned a metallicity drawn at random from the truncated Gaussian distribution appropriate for that galaxy (see Table 2), while ensuring that both members of a binary pair were assigned the same metallicity.

The mass range covered in the draws from an assumed IMF extended both above and below that which is observationally accessible. This is to account for unequal mass-ratio binaries and observational effects. Some stars that are more massive than the

---
[8] Some previous analyses adopted alternative definitions, such as the fraction of sources on the color (apparent) magnitude diagram that are binary pairs (Geha et al. 2013). Denoting this alternative definition by $f_g$, it is straightforward to compare results, using the relation $f_g = \frac{f_f}{2 - f_f}$.

Table 2
Metallicity Information

| Galaxy | $\mu$ | $\sigma$ | Minimum | Maximum | Fiducial Isochrone | Sample Size | Reference |
|---|---|---|---|---|---|---|---|
| Reticulum II | −2.69 | 0.35 | −3.50 | −2.00 | −2.50 | 16 | (1) |
| Ursa Major II | −2.28 | 0.68 | −3.25 | −1.50 | −2.15 | 12 | (2) |
| Triangulum II | −3.08 | 0.27 | −3.50 | −2.50 | −2.50 | 3 | (3) |
| Segue 1 | −2.68 | 0.90 | −3.50 | −1.40 | −2.50 | 7 | (4) |

**Note.** The columns $\mu$ and $\sigma$ give the mean and standard deviation from fitting a Gaussian to the metallicity distribution for each UFD. The next two columns give the minimum and maximum metallicity values used in the creation of synthetic photometric data sets (see Section 3.2; note that the effective minimum metallicity is [Fe/H] = −2.50), while the following columns give the metallicity of the fiducial isochrone adopted to match the observed color–magnitude diagram and the number of stars available to define the distribution, respectively. References: (1) Simon et al. (2015); considering only the set of stars with metallicity estimates from Giraffe spectroscopy, (2) Kirby et al. (2013); note that the two most metal-rich stars ([Fe/H] ∼ −1.0) are plausibly nonmembers, as discussed in Section 3.2.1 ; we chose to truncate the metallicity distribution at a value below these two stars, as described in the text; (3) Buttry et al. (2022); we used only the stars with both membership flag and feh_quality flag equal to unity and averaged the repeat measurements for individual stars; (4) Frebel et al. (2014).

observed regime could have a lower-mass close binary companion that does lie within the targeted mass range, but is "lost" in the unresolved binary. Similarly, stars with masses below the observational limits could be paired into binaries that are observable. Further, photometric errors could scatter some stars to be within the observational limits. We therefore adopted a lower-mass bound of 0.11 $\mathcal{M}_\odot$, which ensures that stellar models exist for all relevant metallicities. The upper-mass limit is less critical, and we chose 1 $\mathcal{M}_\odot$, thus avoiding the need to make assumptions about the form of the IMF at higher masses. The main-sequence turnoff is safely below this maximum mass limit.

We then used stellar isochrones to generate synthetic photometry for each "star" from these assigned physical parameters, interpolating within an isochrone grid of a range of metallicities but fixed age. We next added the fluxes of both members of each binary, and converted the absolute magnitudes of the single "stars" and binary pairs to apparent magnitudes, using the appropriate distance modulus and dust extinction for each galaxy. As in Paper I, we allowed for uncertainties in these quantities by drawing the values of distance and $A_V$ from uniform distributions centered on the values given in Table 1 and with widths of ±1 kpc and ±20%, respectively. Finally, we modeled the effects of photometric error and incompleteness and applied magnitude cuts that were identical to those applied to the observed photometric data (see Table 1).

### 3.2.1. Metallicity Distributions

Ideally, we would follow the approach of Paper I, where we combined [α/Fe] and [Fe/H] information to model the total metallicity distribution ([M/H]) of Boo I. However, this is not possible due to the significantly more limited information on the [α/Fe] distributions of the galaxies in our present sample. Instead, we modeled the metallicity distribution of each UFD galaxy as a truncated Gaussian distribution derived from a fit to published [Fe/H] data. The mean and standard deviation of each resulting fit is given in Table 2, together with the minimum and maximum values that were adopted in the truncation. We prioritized a larger sample over higher spectral resolution when selecting spectroscopic data to use. Further, as discussed in Section 2 above, the minimum metallicity of the isochrone grid we adopted (at [Fe/H] = −2.50) resulted in the very low-metallicity system Tri II being effectively modeled as a single-metallicity population.[9]

---

[9] We did also model the metallicity distribution of Tri II using a different catalog (Kirby et al. 2017) that included higher-metallicity stars and found negligible differences to the results presented in Section 4.2.3.

The metallicity distribution of UMa II that we adopted was based on the analysis of Kirby et al. (2013; itself based on the observations of Simon & Geha 2007). That data set includes two stars of unexpectedly high [Fe/H] for an UFD galaxy (one at [Fe/H] = −1.00, the other at [Fe/H] = −1.11, with the thirdmost metal-rich star at [Fe/H] = −1.86). Further, these two stars are offset from the remainder of the sample in proper-motion space, as shown in Figure A1 in Appendix A. We thus opted to truncate the metal-rich side of the Gaussian metallicity distribution for UMa II at [Fe/H] = −1.50, i.e., between the metallicity of the secondmost and thirdmost metal-rich stars. We verified that there are no such proper-motion-discrepant stars in the catalogs used to define the metallicity distributions of the other UFD galaxies.

### 3.3. Approximate Bayesian Computation Markov Chain Monte Carlo

The galaxies in our sample with a large enough number of observed member sources ($n_{obs} > 600$), namely, Ret II and UMa II, were analyzed with forward modeling of the observed color–magnitude diagram and an ABC-MCMC analysis. The general idea of the ABC-MCMC analysis is to identify the best-fit binary fraction and best-fit IMF parameters for a given galaxy through comparison of the real, observed data—in the form of the normalized Hess diagram—with the corresponding Hess diagrams for many realizations of different simulated, synthetic stellar populations. All Hess diagrams were created using bins of 0.2 mag in both color and apparent magnitude and normalized such that the counts in each Hess diagram summed to unity. In each realization of a population with a given IMF functional form, we drew the values of the binary fraction and IMF parameters from prior distributions that were identical to those we adopted in our earlier analysis of the low-mass IMF in Boo I (see Paper I). We followed the same procedures to implement the ABC-MCMC analysis as in Paper I, with the exception that, in the case of the lognormal IMF, we adopted a final tolerance threshold of 1% rather than the 0.5% used in Paper I, and we ran the MCMC for an additional 4000 iterations after the 1% threshold was met.

Again, as discussed further in Paper I, we adopted flat priors for the parameters of the different forms of the IMF. Specifically, the slopes for both the single power-law and broken power-law IMFs were each drawn from a prior with minimum and maximum bounds of −4.5 and −0.05, respectively (with the implicit understanding that convergence of the total mass would be achieved by the slope below $0.11\mathcal{M}_\odot$, the lower limit of the present analysis).

We verified the ability of the ABC-MCMC analysis to accurately retrieve IMF parameters and binary fractions when applied to "observed" catalogs of only ∼700 synthetic stars. In these verification tests, we applied the ABC-MCMC analysis to synthetic populations with known input IMF parameters and binary fractions, and we compared the output best-fit values and posterior distributions to the input known parameter values (see the Appendix of Paper I for further discussion of these tests). For each synthetic population, we adopted physical parameters (e.g., distance and metallicity distribution) that matched those of UMa II, and we generated synthetic populations with each of the lognormal, broken power-law, and single power-law form of the IMF. For the majority of the tests, the input IMF parameters and binary fractions were within the 68% credible interval (CI) of the posteriors. However, for some combinations of the parameters for the lognormal form of the IMF, the input $\sigma$ parameter was occasionally just outside the lower limit of the 68% CI (by <0.1).[10] The differences between the best-fit values (medians) and the lower limit of the 68% CIs for the $\sigma$ parameter in this subset of tests were smaller than those of the ABC-MCMC results for the observed data. Further, the posteriors in this subset of tests were more biased to high $\sigma$ than the posteriors of both the observed data and the tests for which the input value was within the 68% CI. We thus conclude the true $\sigma$ parameter for the observed UMa II data is likely inside or very nearly inside the 68% CI.

### 3.4. Kolmogorov–Smirnov Tests

For each of the four galaxies of the sample (independent of sample size), we used the two-sample KS test to determine if we could reject the null hypothesis that the sources in a given galaxy's catalog—in the form of their cumulative apparent magnitude distribution in each of the two bandpasses—were drawn from the same distribution as synthetic sources generated with a known IMF and binary fraction. However, as discussed above, there can be rather large sample-to-sample variation between different draws from the same IMF. This variation is amplified by photometric errors, incompleteness, and the random binary pairing, among other effects. We thus opted to generate many synthetic stellar populations, namely, 1000, for each combination of IMF and binary fraction, and computed KS tests between each synthetic population and the real, observed data.

We investigated combinations of IMF parameters and binary fractions on the following grids. For the single power-law IMF, we adopted slope values from −0.5 to −3.5, with a spacing of 0.25 (13 slope values in total). We also adopted this range for both slopes ($\alpha_1$ and $\alpha_2$) of the broken power-law IMF, and considered every possible combination of these values. We fixed the break mass to $M_B = M_{BK} = 0.5\ \mathcal{M}_\odot$. For the lognormal IMF, we adopted values for the characteristic mass ($M_c$) that started at $0.05\ \mathcal{M}_\odot$ and then ranged from $0.1\ \mathcal{M}_\odot$ to $0.9\ \mathcal{M}_\odot$, evenly spaced by $0.1\mathcal{M}_\odot$ (thus 10 values of $M_c$). The values of $\sigma$ for the lognormal IMF started at 0.25 and were evenly spaced by 0.1 from 0.3 to 1.0 (thus nine $\sigma$ values). Again, we considered every possible combination of $M_c$ and $\sigma$.

We also included the values found by Chabrier ($M_{cc} = 0.2\ \mathcal{M}_\odot$, $\sigma_c = 0.55$). Finally, for each combination of IMF parameters for each IMF functional form, we adopted binary fractions of $f_f = 0.1$, 0.5, and 0.8.

We generated 1000 synthetic data sets for each combination of binary fraction and IMF parameters to model the apparent magnitudes in each observed filter. We chose 1000 as a compromise between computational constraints and a statistical number of realizations of the data. Further, this number allowed us to adopt a limit of 1 in 1000 of the synthetic data sets (i.e., 0.1% or between $3\sigma$ and $4\sigma$ in the limit of a one-dimensional Gaussian) in our investigation, as discussed below. With these realizations, it is possible to identify combinations of IMF parameters and binary fractions that can be ruled out. We first estimated the number of masses that needed to be drawn from each IMF combination such that the synthetic data sets would contain approximately as many sources as the real, observed data, once the binary fraction and observational effects (e.g., incompleteness) were incorporated. We then generated the synthetic data and performed two-sample KS tests on the F814W and F606W photometric data separately.[11]

Each KS test was performed with a confidence level of 95%. That is, if the returned $p$-value were less than 0.05 for the F814W and/or the F606W magnitudes, we rejected the null hypothesis and concluded that the data were likely drawn from a population with a different underlying IMF and binary fraction. Should the $p$-value be greater than this threshold ($p > 0.05$) for both F814W and F606W, we considered it to be plausible that the observed data and the synthetic data were drawn from the same distribution. We considered the null hypothesis to be rejected only if *all* of the 1000 realizations with that set of binary fraction and IMF parameters were rejected, since, if even one of the realizations had $p > 0.05$ for both the F814W and F606W filters, there is some possibility that the observed data could have been drawn from the same distribution as the synthetic data. We err on the side of caution in our analysis and adopt this conservative limit to reflect the limited constraining power of the data, especially in light of the unknown levels of contamination from, e.g., background galaxies. A given combination that cannot be rejected is therefore considered plausible, even if it is not very probable (i.e., if only one or a few synthetic data sets have $p > 0.05$ for both filters). We use these tests to identify the combinations that we can confidently rule out, and to map the range of combinations that we cannot. Verification of the reliability of our approach is given in Appendix B, where we also present the results of this method when applied to the Boo I data that we analyzed in Paper I.

### 4. Results

We first present the results from the ABC-MCMC analysis applied to Ret II and UMa II, defining the "best-fit" value of each parameter—including the binary fraction—to be the median of the posterior distribution that we obtained and the CIs to be, respectively, the smallest regions containing 68% and 95% of the posterior probability. We then discuss the outcomes of the KS tests for each of the four galaxies in our sample, identifying the combinations of parameter values that cannot be rejected.

---

[10] We verified that, for the same choices of IMF parameters and binary fraction, the input parameters were within the 68% CI when the ABC-MCMC was applied to a larger synthetic population, with a few thousand stars.

[11] We used the `scipy` function `scipy.stats.kstest` with the "method" parameter set to exact to compute each value, with the asymp method used should the exact method be unsuccessful.

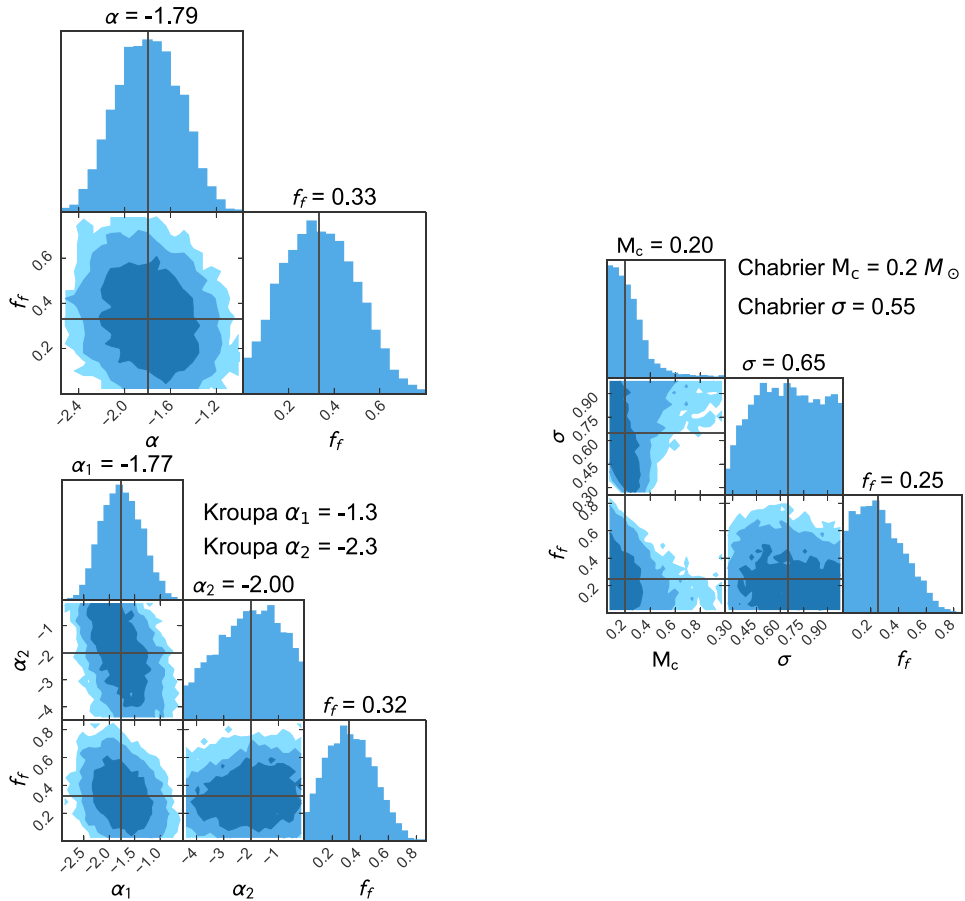

**Figure 4.** The posterior distributions from the ABC-MCMC analysis for Ret II. The results are shown for the parameters of the single power-law (upper left), lognormal (right), and broken power-law (lower left) forms of the IMF. The one-dimensional histograms along the diagonal give the marginalized posteriors. The pairwise projections are plotted beneath the one-dimensional histograms, with the intensity of color indicating the 68%, 95%, and 99% contours (from darkest to lightest shade of blue). The vertical and/or horizontal lines on each panel mark the best-fit values (i.e., the median values of the marginalized posterior distributions), with their numerical values given above the relevant plot.

*4.1. ABC-MCMC Analyses of Reticulum II and Ursa Major II*

We considered three functional forms of the single-star IMF: a single power law, a broken power law (with break mass fixed to the value found by Kroupa 2001, namely, $0.5 \mathcal{M}_\odot$), and a lognormal distribution. We also explored the binary fraction, as defined in Section 3.2 above.

*4.1.1. Reticulum II*

The posterior distributions from the ABC-MCMC analysis for each of the three functional forms of the IMF are shown in the three corner plots in Figure 4. The best-fit (median) values for the parameters are given along the tops of the subpanels of each plot and are indicated by the black lines. The best-fit binary fraction is ~0.3, independent of the adopted form of the IMF. The values found for the stellar population in the solar-neighborhood over this mass range are given for reference.

It is apparent from the distributions shown for the broken power-law form of the IMF (lower left panel of Figure 4) that the slope for lower masses ($\alpha_1$) is more strongly constrained than is the slope for larger masses ($\alpha_2$). This is unsurprising, as there are simply fewer stars available to constrain the fit above the break mass than below it. Further, a higher amplitude sample-to-sample variation is expected at the higher-mass end when relatively small samples are drawn from the IMF (see Figure 3). The best-fit values for the two slopes of the broken power-law IMF are $\alpha_1 = -1.77$ and $\alpha_2 = -2.00$. The Kroupa value for $\alpha_1$ ($\alpha_{1K} = -1.30$) formally lies outside the 68% CI of the posterior, while the Kroupa value for the upper slope ($\alpha_{2K} = -2.30$) lies well inside the 68% CI of the posterior for that slope (see Table 3). That said, the slight tension between the best-fit value for $\alpha_1$ and the Kroupa value should be interpreted with caution: as discussed in Paper I, fixing the value for the break mass, as done here ($M_B = M_{BK} = 0.5\ \mathcal{M}_\odot$), could cause the CIs from the posteriors to be too narrow (see also El-Badry et al. 2017), and wider CIs would likely alleviate the tension. Further, quoted uncertainties in the slopes of the broken power-law IMF found in the Milky Way are ±0.3 for $\alpha_{1K}$ and ±0.36 for $\alpha_{2K}$ (see Equation (55) of Kroupa et al. 2013), which, if taken into account, would reduce the robustness of the tension.

Turning to the case of the lognormal IMF, the right-hand corner plot in Figure 4 demonstrates a covariance between the two IMF parameters, namely, the characteristic mass ($M_c$) and the dimensionless width ($\sigma$)—as has been noted previously (El-Badry et al. 2017)—in the sense that lower $M_c$ values generally correspond to higher $\sigma$ values. The best-fit parameter values are $M_c = 0.20\ \mathcal{M}_\odot$ and $\sigma = 0.65$. The Chabrier values for the Milky Way IMF lie well within the 68% CIs of the posteriors.

The best-fit single power-law slope, indicated in the upper right panel of Figure 4, is $\alpha = -1.79$. While this differs from the familiar Salpeter slope ($\alpha_S = -2.35$), and indeed, the Salpeter value lies outside of the 95% CI of the posterior; a

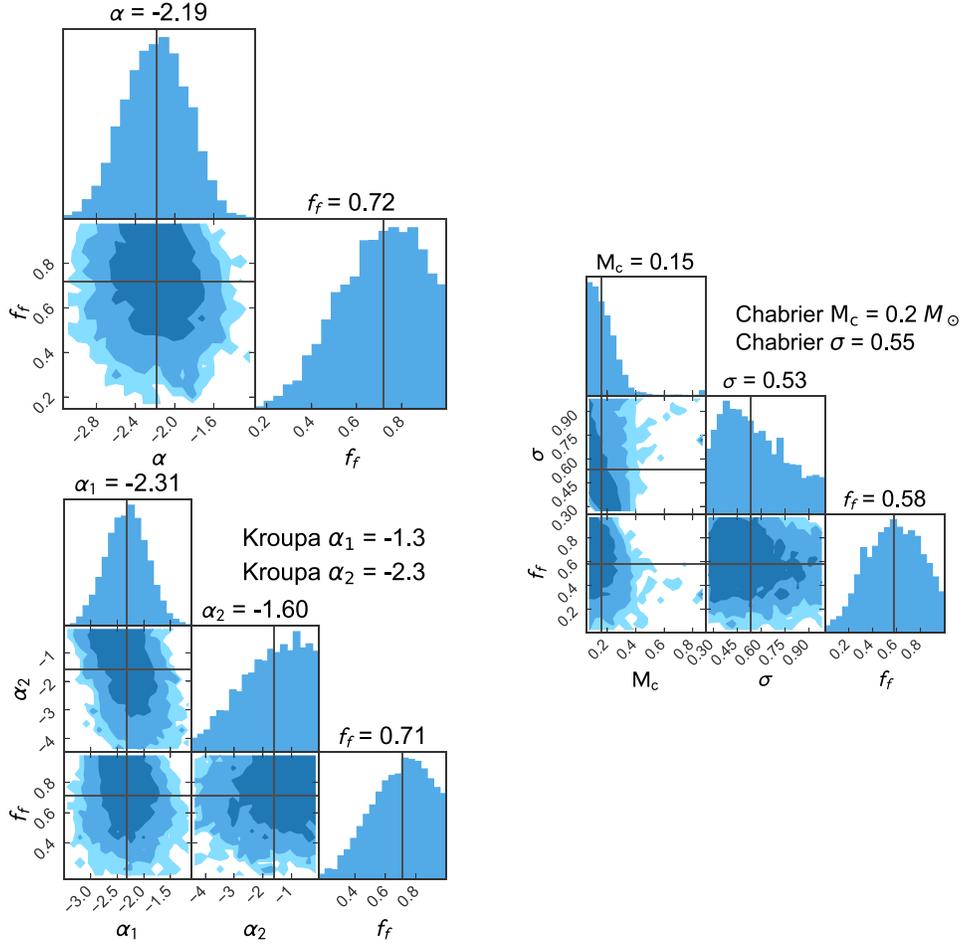

**Figure 5.** The posterior distributions from the ABC-MCMC analysis for UMa II. The results are shown for the parameters of the single power-law (upper left), lognormal (right), and broken power-law (lower left) forms of the IMF. See Figure 4 for details.

Table 3
The Best-fit (Median) IMF Parameters for Ret II and UMa II, Together with the Values that Correspond to the 68% and 95% Credible Intervals (CIs) from the Posterior Distributions

| Galaxy | Single Power Law | | | Broken Power Law | | | Lognormal | | |
|---|---|---|---|---|---|---|---|---|---|
| | Best Fit | 68% CI | 95% CI | Best Fit | 68% CI | 95% CI | Best Fit | 68% CI | 95% CI |
| Ret II | $\alpha = -1.79$ | $-2.06, -1.50$ | $-2.31, -1.29$ | $\alpha_1 = -1.77$ | $-2.15, -1.37$ | $-2.47, -1.04$ | $M_c = 0.20$ | 0.05, 0.26 | 0.05, 0.50 |
| | | | | $\alpha_2 = -2.00$ | $-3.07, -0.63$ | $-4.01, -0.06$ | $\sigma = 0.65$ | 0.45, 0.90 | 0.33, 1.00 |
| | $f_f = 0.33$ | 0.14, 0.47 | 0.03, 0.61 | $f_f = 0.32$ | 0.14, 0.48 | 0.02, 0.62 | $f_f = 0.25$ | 0.01, 0.35 | 0.00, 0.57 |
| UMa II | $\alpha = -2.19$ | $-2.48, -1.83,$ | $-2.79, -1.60$ | $\alpha_1 = -2.31$ | $-2.67, -1.92$ | $-3.08, -1.63$ | $M_c = 0.15$ | 0.05, 0.20 | 0.05, 0.33 |
| | | | | $\alpha_2 = -1.60$ | $-2.24, -0.08$ | $-3.69, -0.05$ | $\sigma = 0.53$ | 0.27, 0.66 | 0.25, 0.93 |
| | $f_f = 0.72$ | 0.60, 0.97 | 0.39, 1.00 | $f_f = 0.71$ | 0.57, 0.96 | 0.37, 1.00 | $f_f = 0.58$ | 0.38, 0.85 | 0.20, 0.99 |

**Notes.** For context, the Salpeter IMF slope is $\alpha_S = -2.35$, the Kroupa single-star IMF has slope values of $\alpha_{1K} = -1.3$ for the lower masses, $\alpha_{2K} = -2.3$ for the upper masses, with the power-law break being at a mass of $M_{BK} = 0.5\ \mathcal{M}_\odot$, and the Chabrier single-star IMF has $M_{cc} = 0.2\ \mathcal{M}_\odot$, and $\sigma_c = 0.55$.
[a] The break mass, $M_B$, was fixed at $M_{BK} = 0.5\ \mathcal{M}_\odot$.
[b] Note that the lower bounds of the CIs for $M_c$ correspond to the lower limit of the prior, which was imposed to ensure that all IMFs populated color–magnitude diagrams within the observational limits.

single power-law IMF with the Salpeter slope is known to overpredict the number of low-mass stars in the Milky Way (e.g., Kroupa 2001; Chabrier 2005).

*4.1.2. Ursa Major II*

The posterior distributions from the ABC-MCMC analysis of the data for UMa II are shown in Figure 5, with the same format as for Ret II. The best-fit values for the two slopes of the broken power-law IMF in UMa II are, respectively, $\alpha_1 = -2.31$ (close to the Salpeter value) and $\alpha_2 = -1.60$. The Kroupa value for the lower-mass segment ($\alpha_{1K} = -1.3$) lies outside of both the 68% CI and 95% CI of the $\alpha_1$ posterior, while the Kroupa value for the higher-mass segment ($\alpha_{2K} = -2.3$) lies well within the 68% CI of the $\alpha_2$ posterior. Interestingly, the best-fit $\alpha_1$ slope is steeper than the $\alpha_2$ slope, such that the best-fit IMF turns up instead of down toward lower masses. This result is surprising in the context of both our understanding of the stellar populations of the Milky Way and results from earlier analyses of the IMF in UFD galaxies, which found either consistency with the Milky Way (Paper I), or indications of a more bottom-light IMF in some systems (see, e.g., Geha et al. 2013; Gennaro et al. 2018a, albeit for the single power-law and/or lognormal forms of the IMF). This upturn at low masses could indicate residual contamination from faint background galaxies, the possibility of which was discussed in Section 2.2.1 above.

The best-fit lognormal IMF parameters are consistent with the Chabrier values, which are encompassed by the 68% CIs of the posteriors of $M_c$ and $\sigma$, albeit that the best-fit $M_c$ value corresponds to a relatively bottom-heavy IMF ($M_c = 0.15$ $\mathcal{M}_\odot$), and the Chabrier value corresponds to the upper edge of the 68% CI. Figure 5 again shows that $M_c$ and $\sigma$ are covariant, with a somewhat tighter trend than for Ret II. Note that values for the characteristic mass that are larger than the best-fit value tend to correspond to width parameter values that are smaller than the best-fit value. This somewhat higher level of covariance suggests that single-value estimates (e.g., the medians and CIs) from the marginalized posteriors may not provide best descriptions of the results.

The best-fit single power-law slope for UMa II is shallower than the Salpeter value ($\alpha = -2.19$ compared to $\alpha_S = -2.35$), although the Salpeter slope lies within the 68% CI of the posterior.

It can be seen that the isochrones overplotted on the data for UMa II in Figure 2 do not provide as good a match to the locus of likely members as for the other three systems. We reanalyzed the data after adopting a higher extinction and found little impact on the results, with the exception of a lower binary fraction.

Thus, the best-fit parameters for all three functional forms of the IMF analyzed formally favor a bottom-heavy IMF in UMa II. However, as discussed above and in Section 2, the possibly nonnegligible contribution from low-luminosity background galaxies (at $z \sim 0.5$) in this field renders this conclusion insecure.

### 4.1.3. Binary Fraction

The best-fit binary fractions that we found for Ret II ($f_f \sim 0.3$) and for UMa II ($f_f \sim 0.6$) are consistent with those determined for other UFD galaxies in similar analyses of the low-mass IMF, which range from $f_f \sim 0.1$ to $\sim 0.76$ (where necessary converting to our definition of binary fraction; see Geha et al. 2013; Gennaro et al. 2018a, 2018b; and Paper I). In all these determinations, the best-fit binary fraction is constrained with photometric data and thus is largely determined by the width of the color–magnitude diagram.

Ret II is the sole galaxy for which we can compare the best-fit binary fraction from the ABC-MCMC analysis to independent estimates, albeit derived using different techniques.[12] There are two published results, the first being based on multiepoch measurements of line-of-sight velocities of red giant stars (Minor et al. 2019), while the second focuses on wide binaries and is based on a search for resolved pairs in HST imaging data (Safarzadeh et al. 2022).

It is difficult to make a direct comparison between the results of the present photometry-based analysis for the binary fraction of unevolved stars in Ret II ($\sim 0.3$) and that of the spectroscopy-based analysis of Minor et al. (2019) for red giant stars ($\geqslant 0.5$). Even apart from the uncertainties of binary evolution, and the different sensitivities of the two techniques to parameters such as binary separation and period, the two analyses make different assumptions—for example, modeling the mass ratios—in the creation of the input synthetic binary population. Despite these differences and uncertainties, our results are generally consistent with those of Minor et al. (2019), as a binary fraction of 0.5 lies just outside the 68% CI for the posterior of each of the single power-law and broken power-law IMF, and even higher binary fractions lie inside the 95% CIs for all three forms of the IMF.

The second independent analysis (Safarzadeh et al. 2022) finds a wide-binary fraction of $\sim 0.007$, which they extrapolate to a total binary fraction of $\sim 0.5$ under the assumption that the binary population of Ret II has a distribution of separations similar to that of binary systems of all metallicities in the local Milky Way. Echoing the discussion above (and keeping in mind that there may well be metallicity-dependent systematics in the distribution of separations, as found by, e.g., Moe et al. 2019; Hwang et al. 2021), we conclude that our binary fraction results are also generally consistent with those of Safarzadeh et al. (2022).

### 4.1.4. Comparisons of Best-fit IMF with Luminosity Function Data

Section 4.2 below presents the results of the KS tests between the observed data in the form of empirical cumulative luminosity functions and the predictions for synthetic populations created using a grid of IMF parameter values and binary fractions. In this spirit, we compared the observed cumulative luminosity functions for each of Ret II and UMa II with the predictions for 100 realizations of synthetic populations created with parameter values drawn from the posterior distributions from the ABC-MCMC analysis of the full color–magnitude data. We present these unbinned, cumulative luminosity functions in Figure 6. At the faintest magnitudes, the cumulative luminosity functions of the observations are below those of the synthetic data, as would be expected if there is increased contamination from, e.g., background galaxies at fainter magnitudes. Over the vast majority of the magnitude range, however, there is good agreement between the observations and the synthetic data.

### 4.2. KS Tests of the Full Sample of Four Galaxies

The results of the KS tests are presented in the form of plots showing the two-dimensional grids of the IMF parameters that were tested. These are the binary fraction and slope, for the single power-law IMF, the two slopes for the broken power-law IMF—with different values of the binary fraction indicated by the symbol used—and the characteristic mass and the width parameter for the lognormal IMF, again with different values of the binary fraction indicated by different symbols. Grid points at combinations of given parameter values that are *not* rejected by the KS tests (i.e., with $p > 0.05$ for both filters for at least one of the 1000 synthetic data sets) are shown with green symbols.

---

[12] There is also a published investigation of the binary population of Seg 1 based on measurements of the line-of-sight velocity distribution (Martinez et al. 2011), but we did not determine a best-fit binary fraction for Seg 1.

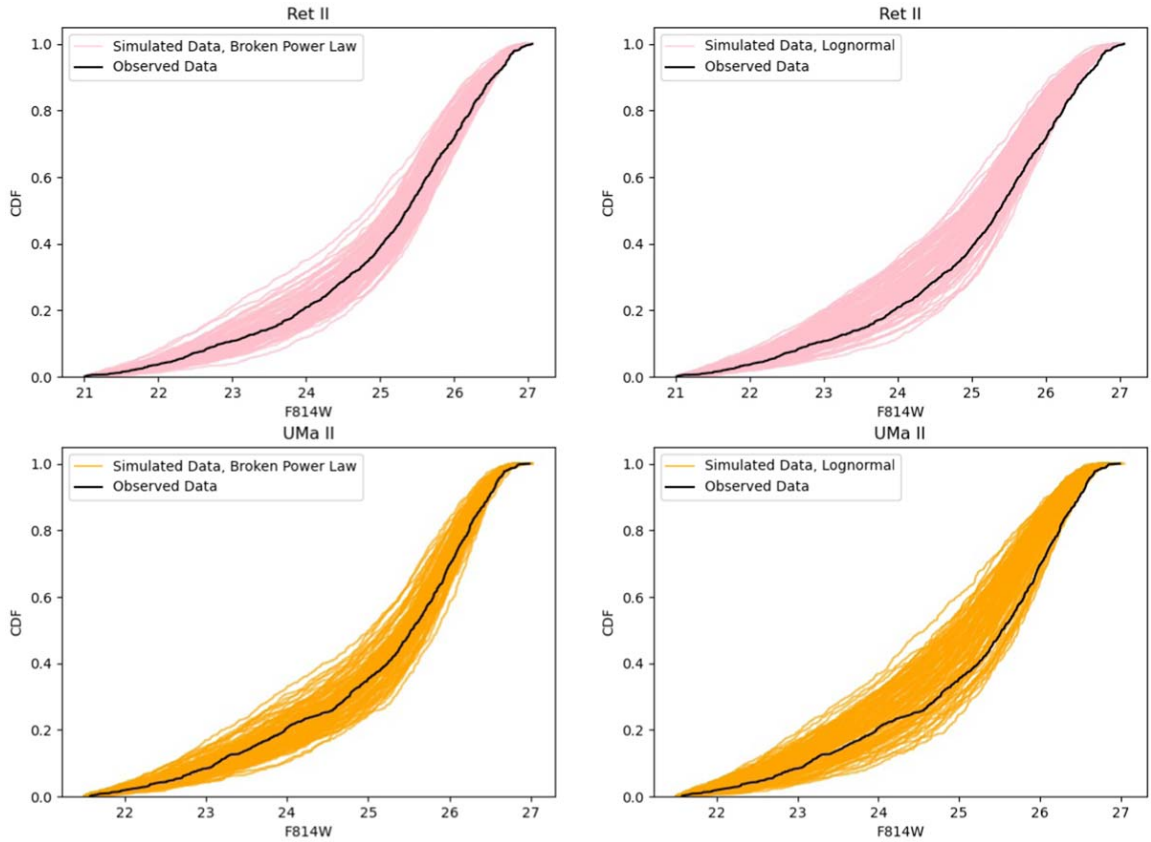

**Figure 6.** Upper panels: Comparisons between the observed data for Ret II, in the form of the cumulative luminosity function in the F814W bandpass (solid black curves), with the predicted functions (pink curves) for 100 realizations of synthetic stellar populations with IMF parameters drawn from the posterior distributions obtained in Section 4.1.1. The left upper panel shows predictions for the broken power-law IMF, while the right upper panel shows predictions for the lognormal IMF. Lower panels: as the upper panels, but for UMa II, using the posterior distributions from Section 4.1.2 (orange curves).

### 4.2.1. Reticulum II

The results of the KS tests for Ret II are presented in Figure 7. While the binary fractions explored in these KS tests ($f_f = 0.1$, 0.5, and 0.8) differ from the best-fit value found above ($f_f \sim 0.3$), the results of the KS tests are generally consistent with the results of the ABC-MCMC.

First, for the single power law, slopes from $\alpha = -1.25$ through $-2.25$ are not rejected. For the broken power-law IMF, the KS tests reject values of the $\alpha_1$ slopes that are either shallower than $-1.00$ or steeper than $-2.5$, while for each $\alpha_2$ grid point there exists some combination of the remaining parameters ($\alpha_1$ and $f_f$) that is not rejected. In particular, the grid point closest to the values for the slopes found by Kroupa ($\alpha_1 = -1.25$, $\alpha_2 = -2.25$) is not rejected. For the lognormal IMF, all combinations of parameters with the characteristic mass at, or above, $0.3\mathcal{M}_\odot$ are rejected. The values for characteristic mass and width found by Chabrier ($0.2\mathcal{M}_\odot$ and 0.55, respectively) are not rejected, provided the binary fraction equals 0.1. We find that combinations of low $\sigma$ and $M_c$ are rejected, while combinations of higher $\sigma$ and low $M_c$ are not, which mimics the posterior of the ABC-MCMC results above.

### 4.2.2. Ursa Major II

The results of the KS tests for UMa II are shown in Figure 8. As was the case with Ret II, the results of the KS tests and the ABC-MCMC analyses are generally consistent.

Values that are not rejected for the slope of a single power-law IMF lie in the range $-1.50 \geqslant \alpha \geqslant -2.75$, which encompasses the value found by Salpeter ($\alpha = -2.35$). In the case of the broken power-law IMF, values of the lower-mass slope ($\alpha_1$) that are in the range $-1.25 \leqslant \alpha_1 \leqslant -3.25$ are not rejected. For each value of the upper-mass slope ($\alpha_2$) tested, there exists some combination of $\alpha_1$ and binary fraction ($f_f$) that is not rejected. Unlike for Ret II, in this case, the grid point closest to the values for the slopes found by Kroupa ($\alpha_1 = -1.25$, $\alpha_2 = -2.25$) is rejected, although the two adjacent points at $(-1.25, -2.5)$ and $(-1.5, -2.25)$ cannot be rejected. As noted earlier, the reported uncertainties in the slopes of the broken power-law IMF found in the Milky Way are of order $\pm 0.3$ (Kroupa et al. 2013), and values within this range are not rejected.

For the lognormal IMF, the only combinations of IMF parameters that are not rejected have $M_c < 0.4\,\mathcal{M}_\odot$ and $\sigma < 0.9$. The values found by Chabrier ($0.2\mathcal{M}_\odot$ and 0.55) are formally rejected, but grid points at slightly lower values of $\sigma$ and/or lower values of $M_c$ are not rejected, e.g., $(M_c, \sigma) = (0.2\mathcal{M}_\odot, 0.5)$, provided the binary fraction is low, $\sim 0.1$. Unfortunately, there are no uncertainties given for the IMF parameter values in Chabrier (2005), although the earlier analysis in Chabrier (2003) quotes $\sim 0.02\,\mathcal{M}_\odot$ in $M_c$ and $\sim 0.04$ in $\sigma$. It can be seen that viable combinations match increasingly smaller values of $\sigma$ with larger values of $M_c$, echoing the trends found by the ABC-MCMC analysis (shown in Figure 5).

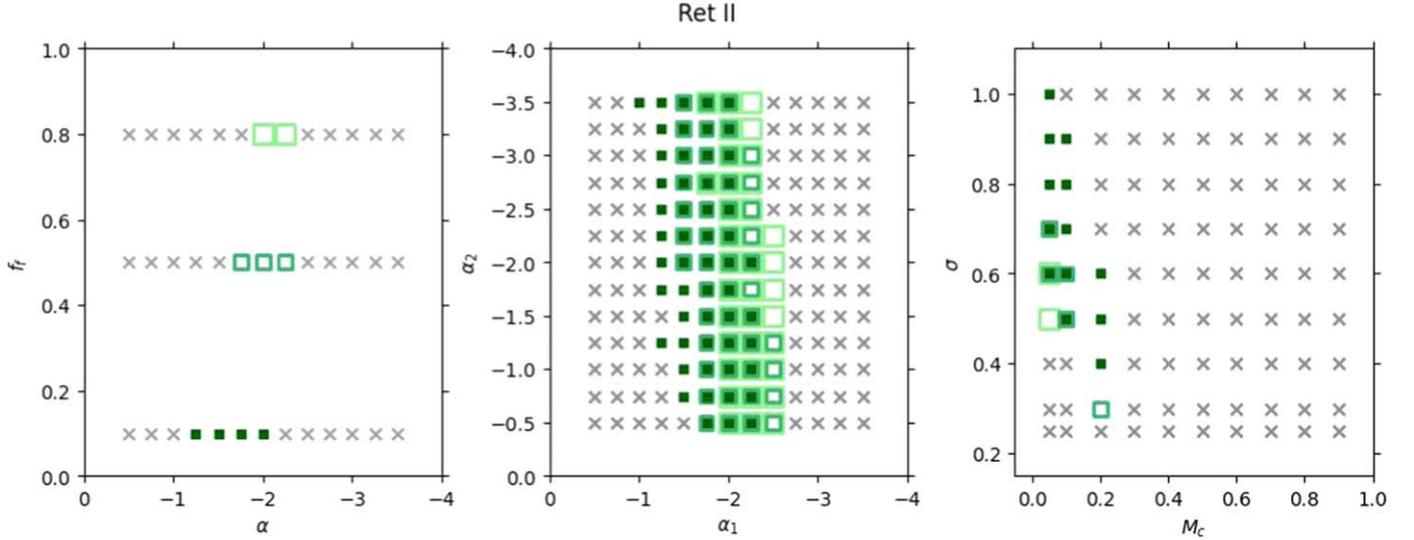

**Figure 7.** The results of the KS tests for Ret II, for grids of the parameters of each of the single power-law (left), broken power-law (middle), and lognormal (right) forms of the low-mass IMF. Each point on the plots corresponds to a combination of IMF parameters plus a binary fraction. We investigated binary fractions of $f_f = 0.1$, 0.5, and 0.8, which are shown with small filled green squares, open medium-size green squares, and large open green squares, respectively. A grid point shown with any one of these (green) markers indicates that that combination of binary fraction and IMF parameter values is not rejected (i.e., the $p$-value from the KS test is above 0.05 for the apparent magnitude distributions in both the F606W and F814W filters, for at least one of the 1000 realizations of the corresponding synthetic stellar population). Grid points marked with gray crosses indicated rejected combinations. For the single power-law case (left), there is only one IMF variable, and thus, the $y$-axis corresponds to the binary fraction. We also tested the lognormal IMF with the values $M_{cc} = 0.2\,\mathcal{M}_\odot$, $\sigma_c = 0.55$ reported by Chabrier (2005), as discussed in the text.

### 4.2.3. Tri II

The results of the KS tests for Tri II are shown in Figure 9, where it may be seen that, in the case of the single power law, slopes ranging from $\alpha = -1.5$ to $\alpha = -2.5$ are not rejected, including the grid point closest to the Salpeter slope. For the broken power-law IMF, a broad range of combinations of values of the two slopes ($\alpha_1$ and $\alpha_2$) and the binary fraction ($f_f$) are not rejected, including at least one set for each value of the $\alpha_2$ slope explored. The values that are not rejected for the lower-mass slope span $-1.0 \geqslant \alpha_1 \geqslant -3.0$, and the grid point closest to the Kroupa values is not rejected. In the case of the lognormal IMF, all values of the characteristic mass greater than $0.3\,\mathcal{M}_\odot$ are rejected, while the pair of parameter values found by Chabrier are not rejected, provided the binary fraction equals 0.1. As noted earlier, in Section 3.2.1, tests carried out using mock populations with a finite metallicity spread, extending to higher values, reached the same conclusions (i.e., that a Milky Way–like IMF cannot be rejected) as these KS tests, with no metallicity spread.

### 4.2.4. Seg 1

The results of the KS tests for Seg 1 are shown in Figure 10. Values of the slope of the single power-law IMF ($\alpha$) that are not rejected range from $-1.5$ to $-2.25$, this lowest value being the closest grid point to the Salpeter slope. For the broken power-law IMF, slopes of the lower-mass segment that are shallower than $\alpha_1 = -1.25$ are rejected, as are those that are steeper than $-2.75$. Once again, there exists some combination of the lower-mass slope and binary fraction for each value of the slope of the higher-mass segment ($\alpha_2$) that is not rejected, and again, the grid point closest to the Kroupa values is not rejected. For the lognormal IMF, all values of the characteristic mass that are higher than $0.2\,\mathcal{M}_\odot$ are rejected. The values for the characteristic mass and width reported by Chabrier (2005) are formally rejected, although very similar values are not rejected (each of the following pairs is not rejected: ($M_c$, $\sigma$) = ($0.2\mathcal{M}_\odot$, 0.5), ($0.1\mathcal{M}_\odot$, 0.5), ($0.1\mathcal{M}_\odot$, 0.6)).

### 4.2.5. Synthesis of KS Test Results

An inspection of Figure 7 through 10 shows that a wide range of parameters are rejected, and that a similar range of parameter space cannot be rejected for the parent population in each of the four galaxies. Further, there is a degeneracy between the values of the IMF slopes and binary fraction, for the power-law IMF, and between the characteristic mass and binary fraction for the lognormal IMF. Such a degeneracy has been noted previously (see, e.g., Kroupa et al. 1991; and the discussion in Geha et al. 2013) and likely is behind the trend seen in the plots, such that relatively bottom-light IMFs are not rejected when paired with only the low binary fraction ($f_f = 0.1$), whereas less bottom-light IMFs are not rejected for a wider range of binary fractions.

The exact regions of parameter space that are not rejected for each form of the IMF differ among the galaxies in the sample. However, the significant overlap between these regions indicates that it is plausible that all of these galaxies are populated by the same IMF. This is illustrated in Figure 11, which shows the set of parameter values that are not rejected for the broken power-law and lognormal IMFs that are in common across the sample. It can be seen that, for the majority of galaxies in the sample, a range of both Milky Way–like and non–Milky Way–like IMF parameter values cannot be rejected. The KS test results for simulated stellar populations drawn from either the Kroupa and Chabrier IMFs that are presented in Figure B1 in Appendix B show similar results: even when the underlying IMF is fixed to that of the Milky Way, a range of different IMFs cannot be rejected. As such, the wide range of IMF parameters that are not rejected is not necessarily a sign that the underlying IMF is actually far from that of the Milky

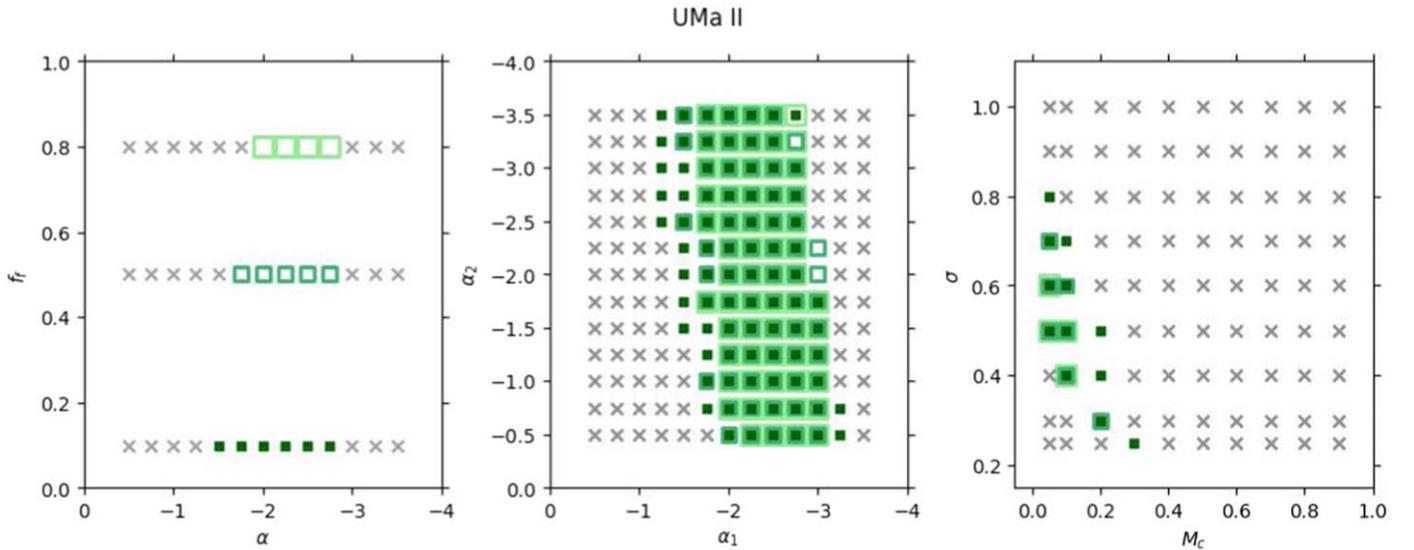

**Figure 8.** The results of the KS tests for UMa II, showing the grids of IMF parameters for the single power-law (left), broken power-law (middle), and lognormal (right) forms of the IMF. The color-coding and symbols follow the scheme of Figure 7.

Way, and could instead be more indicative of the sample-to-sample variation that occurs for small data sets.

A summary of the results from the KS tests between data for each of the UFD galaxies and synthetic populations with assumed IMFs that match those found for stellar populations in the Milky Way is given in Table 4. We stress again here that we employ very conservative limits, and consider a combination of IMF parameters and binary fraction to be rejected only if all of the 1000 realizations have $p \leqslant 0.05$ for one or both photometric filters. As stated in the table, the stellar populations in each of Ret II, Tri II and Seg 1 could have a parent population with an IMF that matches, or is very similar to, that of the Milky Way. The exact reported parameter values for the Milky Way IMF are formally rejected for UMa II, although, as discussed above, a broken power-law IMF with parameter values within the quoted uncertainties of the accepted values for the Milky Way is not rejected, and a lognormal IMF with very similar parameter values to the values for the Milky Way is also not rejected.

## 5. Discussion

There are two main aspects of the low-mass IMF that can be addressed with our sample. The first is whether or not the IMF of metal-poor, ancient populations that formed in low-mass dark-matter haloes, as represented by UFD galaxies, can be distinguished from that of the Milky Way (including the metal-rich, star-forming disk). The second is whether or not it is possible to discuss "the IMF" of all UFD galaxies, and, if not, whether the variations between galaxies show any systematics with galaxy-scale properties.

### 5.1. Comparison with the Milky Way Single-star Low-mass IMF

The best-fit parameter values for the lognormal IMF found for both Ret II and UMa II are consistent with those reported for the Milky Way by Chabrier (2005). In the case of the broken power-law IMF, the best-fit value for the slope at higher masses for each of these two galaxies is consistent with that reported for the Milky Way by Kroupa (2001), while the best-fit lower-mass slope is steeper than for the Milky Way, with the discrepancy being more significant for UMa II. As discussed earlier, taking account of uncertainties in the values for the Milky Way would lessen the tension. Further, there is the possibility of higher residual contamination by faint galaxies in the photometric data of UMa II. We investigate the effect of increasing contamination on the results of the KS tests in Section B.1.4.

As summarized in Table 4, either IMF parameter values that match those reported for the Milky Way, or neighboring values on the grid that was tested, cannot be rejected (i.e., $p > 0.05$ for both photometric filters for at least one of the 1000 synthetic data sets of a given combination of IMF parameters and binary fraction) for each of the entire sample of UFD galaxies. A large number of IMF parameters can be rejected, including many that are very dissimilar to that of the Milky Way.

There is some theoretical expectation that the IMF should depend on metallicity (see, for example, the discussion in Kroupa et al. 2013 and references therein), and indeed, the results of some analyses have suggested that the IMF is metallicity dependent. For example, Li et al. (2023) used both Gaia and LAMOST (spectroscopic) data for local field M-dwarf stars and found that, when they fit single power-law forms of the IMF to metallicity-dependent subsets of the data, the best-fit slope varied systematically with metallicity (note that their approach differed significantly from ours and involves different modeling steps). The expectation that the IMF should depend on metallicity is generally based on models that assume that the Jeans mass sets the relevant scale for stellar masses, so that cooling, largely through metal lines, is essential, even at very low levels of chemical enrichment (e.g., Bromm & Larson 2004). However, the process of fragmentation is complicated, for example, by the possible role of (competitive) accretion (e.g., Bonnell et al. 2001) and/or the formation of binary (and higher multiplicity) systems (e.g., Offner et al. 2023), and it may be that many processes combine to provide an invariant mass function (e.g., the central limit theorem argument of Adams & Fatuzzo 1996, which would favor a lognormal form).

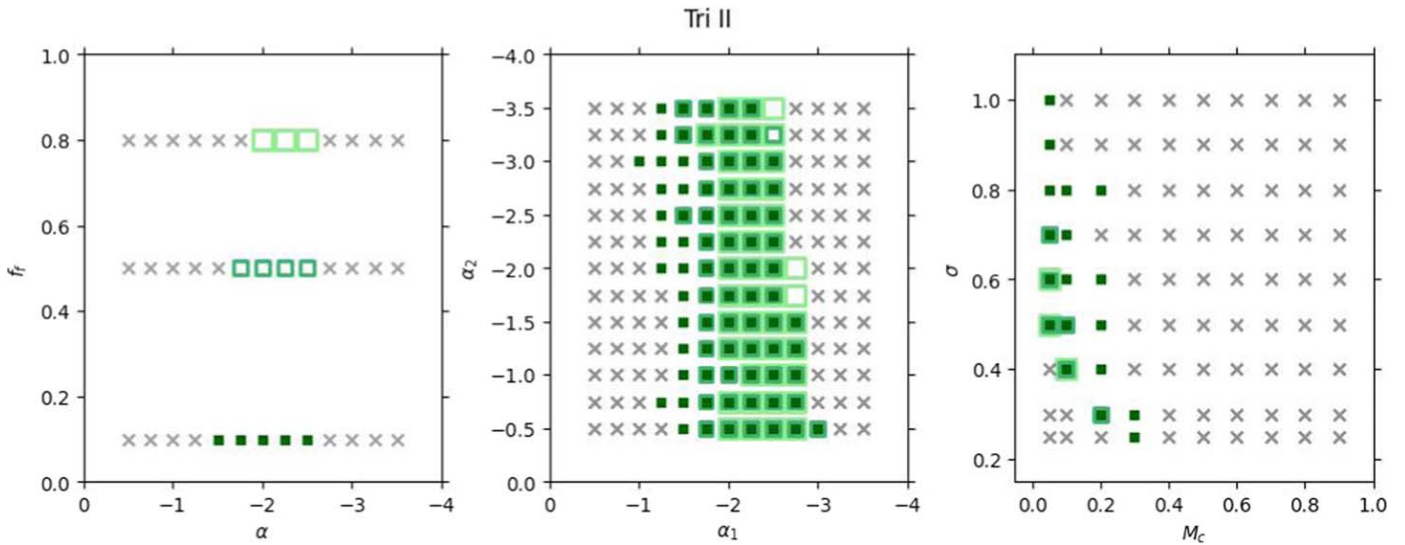

**Figure 9.** The results of the KS tests for Tri II, showing the grids of IMF parameters for the single power-law (left), broken power-law (middle), and lognormal (right) forms of the IMF. The color-coding and symbols follow the scheme of Figure 7.

Our analysis of low-metallicity stellar systems (where the metallicity measurements are from F/G stars) points instead to a plausible overall consistency between the underlying single-star, low-mass IMF in each of the UFD galaxies and that in the Milky Way. Such a similarity is consistent with recent high-resolution simulations of star formation, such as the results of Tanvir & Krumholz (2024), who found only a weak dependence of the low-mass IMF on metallicity, over the range −2 to +0.5 dex. In addition, Alexander et al. (2023) model the chemical evolution of Ret II assuming inhomogeneous enrichment from a Kroupa IMF, and they find that they are able to reproduce many of the observed chemical abundance trends of Ret II members, after tuning of the other input parameters such as star formation efficiency. These results, together with those discussed in the review of Bastian et al. (2010), suggest that the IMF is not sensitive to metallicity (at least above the level of −2.5 dex of the present study). As noted in Section 1, the IMF is essential to many areas of astrophysics, and our analysis indicates that, for low baryonic mass, low-metallicity dark-matter dominated galaxies, the assumption of the typical Milky Way (low-mass) IMF should generally be valid. Note, however, that a range of IMF parameters are consistent with the UFD galaxies in this sample, and thus, the impact of assuming a range of different IMF parameters (including the typical Milky Way parameters) in a given analysis should also be explored.

### 5.2. Galaxy to Galaxy Comparison

As discussed above, and illustrated in Figure 11, there is a large region of overlap in the values of the IMF parameters that are not rejected by the KS tests. It is thus plausible that all of the galaxies formed stars with the same underlying low-mass IMF, albeit that the exact form of that IMF remains uncertain. In this context, the similarity of the value of the so-called Type II plateau in the [$\alpha$/Fe] elemental abundance ratio also argues (Wyse 1998) for an invariant IMF, in this case for massive stars, $\gtrsim 8\ \mathcal{M}_\odot$, as recently demonstrated for UFD galaxies by Reichert et al. (2020).

The mean metallicity varies little across the sample, while the luminosity varies[13] by a factor of ∼15 (see Tables 1 and 2). Any conclusion about the lack of dependence of the IMF on luminosity within this set of UFD galaxies is complicated by the limitations of the small sample sizes for the two lower-luminosity galaxies and the resulting uncertainties (refer back to Figure 3). Similarly, while considering only published radial velocities (and fitting a Gaussian to the distribution) would lead to an estimated factor of $\gtrsim 2$ variation in velocity dispersion (e.g., the compilation in Table 1 of McConnachie & Venn 2020), concerns about contamination by nonmembers make this uncertain (e.g., Table 5 of McConnachie & Venn 2020; and Figure A1 and the associated discussion below).

The fact that the results for Ret II are similar to those for the other galaxies is also interesting since it is likely that Ret II is a satellite of the Large Magellanic Cloud (e.g., Patel et al. 2020), while the rest of the sample are likely satellites of the Milky Way. The chemistry of Ret II member stars is also known to be distinct: a large fraction of Ret II stars shows a significant enhancement in r-process elements (not included in the chemical evolution model of Alexander et al. 2023), while member stars in most UFD galaxies have low r-process abundances (see, e.g., Ji et al. 2016). Detailed studies of more satellite galaxies of the LMC would obviously be of interest, and a systematic survey down to low masses in each of the likely LMC satellites (such as those listed in Patel et al. 2020) would be particularly insightful.

Thus, our analysis finds no evidence for systematic variation of the IMF across the present sample.

### 5.3. Comparison with Results for UFD Galaxies in the Literature

Geha et al. (2013), Gennaro et al. (2018a, 2018b), and Safarzadeh et al. (2022) have each used HST imaging to constrain the low-mass IMF in UFD galaxies. The mass ranges probed for the galaxies in these analyses differ, as do the sizes of the photometric catalogs of likely member sources, with the smallest catalog (containing ∼700 sources) being that for Coma Berenices, analyzed in Gennaro et al. (2018b). Further, while each of those analyses forward-model synthetic stellar populations for comparison to the observed data, the details of the comparisons and methodologies vary, and some analyses aim to constrain the system IMF (including binaries as single sources), rather than the underlying single-star IMF. Direct comparison with the results of the present analysis is therefore difficult, so we largely restrict the discussion to whether or not there is agreement with earlier analyses in terms of comparisons with the Milky Way. As noted in the Introduction, the only overlap in sample with the present work is the analysis of Ret II by Safarzadeh et al. (2022).

---

[13] The long-established relation between luminosity and mean metallicity found at higher luminosities (Kirby et al. 2013) weakens considerably for fainter dwarf galaxies, albeit the internal spread increases (Norris et al. 2010).

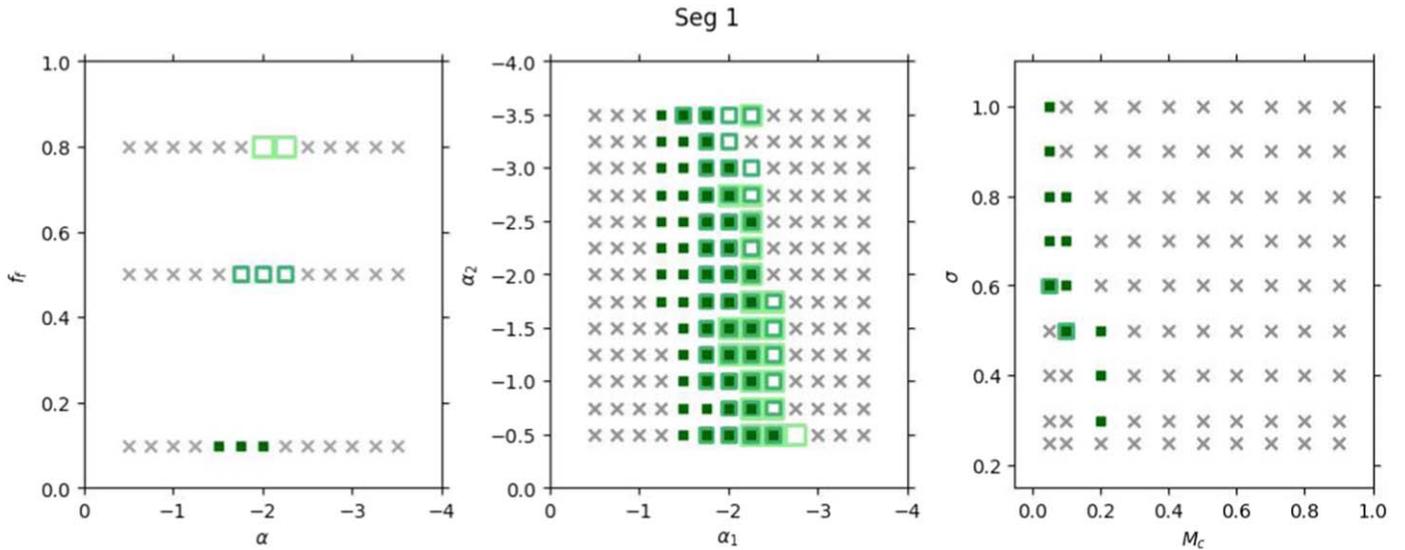

**Figure 10.** The results of the KS tests for Seg 1, showing the grids of IMF parameters for the single power-law (left), broken power-law (middle), and lognormal (right) forms of the IMF. The color-coding and symbols follow the scheme of Figure 7.

First, we present a brief summary of the forms of the IMF and mass ranges probed by these previous analyses. Safarzadeh et al. (2022) analyzed data for only Ret II and constrained only the single power-law form of the IMF, over the mass range $0.34\,\mathcal{M}_\odot$–$0.78\,\mathcal{M}_\odot$. Geha et al. (2013) and Gennaro et al. (2018a) each constrained the single power-law and lognormal forms of the IMF,[14] with Geha et al. (2013) probing the mass range ∼0.5–0.8 $\mathcal{M}_\odot$ and Gennaro et al. (2018a) probing ∼0.4 $\mathcal{M}_\odot$ to ∼0.8 $\mathcal{M}_\odot$. Geha et al. (2013) fixed the lognormal $\sigma$ parameter in their analysis, while Gennaro et al. (2018a) did not. Finally, Gennaro et al. (2018b) and our analysis in Paper I both constrained each of the single power-law, the broken power-law, and the lognormal forms of the IMF, and in Paper I, we fixed the break mass of the broken power-law form. These last two analyses reach down to ∼0.2 $\mathcal{M}_\odot$ and ∼0.3 $\mathcal{M}_\odot$, respectively.

Regardless of whether they constrained the single-star IMF or the system IMF, each of the above analyses found a value for the slope of a modeled single power-law IMF that was shallower than the Salpeter value (i.e., $\alpha > \alpha_S = -2.35$), for every UFD galaxy analyzed. Similarly, the set of single power-law slopes for each galaxy in the present analysis includes those shallower than the Salpeter value. The single power-law slopes found for a mix of single-star and system IMFs in Geha et al. (2013), Gennaro et al. (2018a, 2018b), Safarzadeh et al. (2022), and for Ret II in the present analysis are all shallower than $-2$. The interpretation of these slopes is not straightforward, especially as the Milky Way is known to not be well described by a single power law with a Salpeter slope at low masses. Indeed, as discussed in El-Badry et al. (2017), a shallow single power-law slope is not necessarily indicative of a non–Milky Way–like IMF. Those authors also note that the variation between the best-fit single power-law slope values found in different analyses could be due to the different mass limits of each sample.

The HST data that Safarzadeh et al. (2022) analyze are in the same bandpasses as the present analysis, but are significantly shallower (exposure times ∼1000 s in each filter compared to ≳4600 s) and cover a significantly wider area (a mosaic of 12 ACS fields compared to a single ACS field) as they were obtained primarily to determine the star formation history of Ret II (Simon et al. 2023). The main science goal of Safarzadeh et al. (2022) was an investigation of the wide-binary fraction (discussed in Section 4.1.3 above), and there are several differences between their approach to constraining the IMF and ours. Perhaps the most important difference concerns the treatment of binaries; Safarzadeh et al. (2022) draw primary stars from an assumed IMF, then generate secondary stars according to an assumed mass-ratio distribution, which does not preserve the IMF. In contrast, our approach, explained in detail above, randomly pairs two masses, each of which is drawn from the assumed IMF, thus preserving the IMF. As noted earlier, the lower-mass limit of their analysis ($0.34\,\mathcal{M}_\odot$) is significantly higher than ours ($\lesssim 0.2\,\mathcal{M}_\odot$), which can lead to a different slope being found for a single power-law IMF when fit to an underlying IMF that is not in fact best described by a single power law (El-Badry et al. 2017). With these differences in mind, we note that the best-fit slope that we found from the ABC-MCMC analysis for Ret II is steeper than their "best" range from KS tests on the F814W data. Further, while their

---

[14] Note that Dib (2022) found that the system IMF single power-law slopes in Gennaro et al. (2018a) were consistent with the predictions over the mass range 0.4–0.8 $\mathcal{M}_\odot$ for a model of an integrated galaxy-wide IMF (i.e., the sum of the IMFs in each of the individual star-forming clusters) that includes some variation of the IMF in individual star-forming clusters.

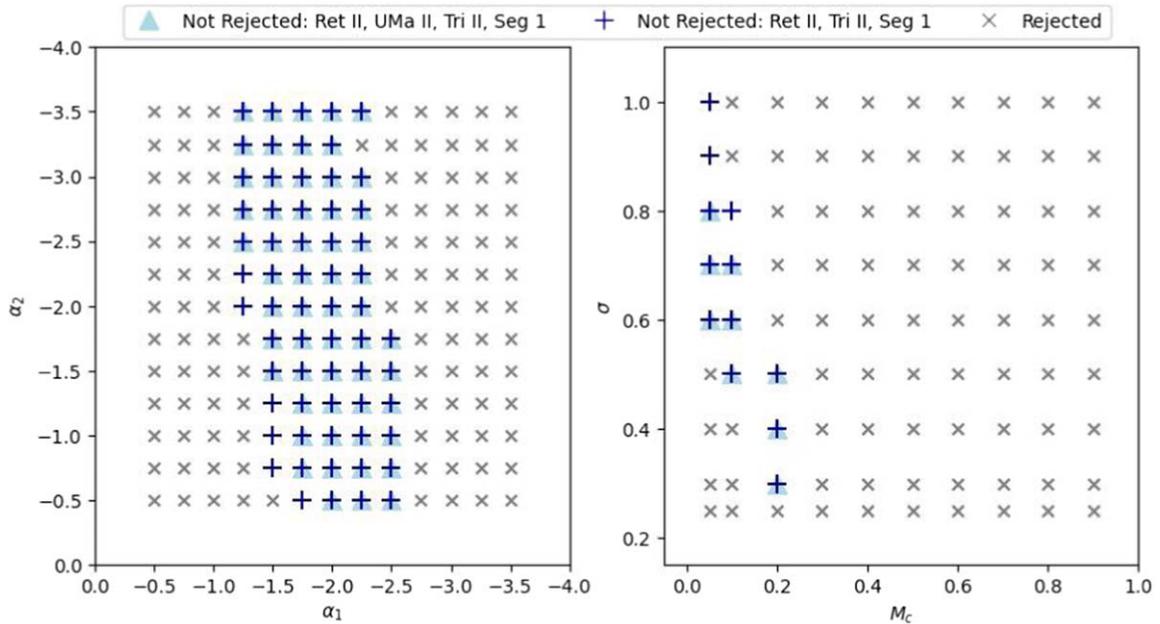

**Figure 11.** The results of the KS tests for the broken power-law (left) and lognormal (right) IMF parameters. Points shown with blue triangles indicate combinations of IMF parameters that are not rejected by the KS tests for the set of of Ret II, UMa II, Tri II, and Seg 1, while blue plus signs show those parameters not rejected by the set of Ret II, Tri II, and Seg 1 (i.e., excluding UMa II). In this, we require only that a minimum of one of the binary fractions investigated be not rejected. The gray x-shaped markers indicate combinations that are rejected by at least one of Ret II, Tri II, or Seg 1.

**Table 4**
Summary of the Results of the KS Tests for the Three Forms of the IMF

| Galaxy | Single Power-law IMF $\alpha = -2.25$ Rejected? | Broken Power-law IMF $\alpha_1 = -1.25$, $\alpha_2 = -2.25$ Rejected? | Lognormal IMF $M_c = 0.2$ $\mathcal{M}_\odot$, $\sigma = 0.5$ Rejected? |
|---|---|---|---|
| Ret II | No | No | No |
| UMa II | No | Yes[a] | No[b] |
| Tri II | No | No | No |
| Seg 1 | No | No | No[b] |

**Note.** The results in this table are those for the synthetic populations with IMF parameter values at the grid points closest to those values reported for stellar populations in the Milky Way. A "No" entry indicates that the specified IMF parameter values are not rejected for at least one of the 1000 realizations of the synthetic population, for at least one choice of binary fraction. A "Yes" entry indicates that the entire suite of realizations is rejected for all three values of the binary fraction considered (0.1, 0.5, or 0.8).
[a] The neighboring grid points are not rejected.
[b] The values $M_{cc} = 0.2$ $\mathcal{M}_\odot$, $\sigma_c = 0.55$ reported by Chabrier (2005) are rejected.

"best" range does not overlap with the 95% CI of our Ret II ABC-MCMC posterior, their 1σ range does. This overlap indicates that their results may not be in tension with our own, despite the differing choices made in the analyses.

The results of our analysis of the broken power-law IMF for Ret II, Tri II, and Seg 1 are consistent with the results of the corresponding analyses presented in Gennaro et al. (2018b) and in Paper I. Namely, the data for Ret II, Tri II, and Seg 1 and the galaxies investigated in Gennaro et al. (2018b) and Paper I (Coma Berenices and Boo I, respectively) do not reject the possibility that the underlying IMF matches that of the Milky Way. While we cannot directly compare our best-fit values for the IMF slopes with those of Gennaro et al. (2018b), who constrain the system IMF, we can compare our values with those we derived in Paper I for Boo I. The best-fit values for the lower-mass slope for Ret II and Boo I ($\alpha_1 = -1.77$ and $\alpha_1 = -1.67$, respectively) are steeper than the value found by Kroupa. Taken at face value, this steeper slope would indicate that there are relatively *more* stars below 0.5 $\mathcal{M}_\odot$ in Ret II and Boo I than in the Milky Way. However, the uncertainties in the best-fit values in both analyses are large, as are the uncertainties in the Kroupa values themselves (±0.3 for $\alpha_{1K}$ and ±0.36 for $\alpha_{2K}$, see Kroupa et al. 2013). The results found above for UMa II are, at face value, not consistent with those in Gennaro et al. (2018b) or in Paper I.

The results of the analysis of the lognormal IMF for Ret II and Tri II that we found above are consistent with the results (for different UFD galaxies) of Geha et al. (2013), Gennaro et al. (2018b), and Paper I, who each find best-fit values for the parameters of the lognormal IMF that are consistent with those of the low-mass stellar populations of the Milky Way. The comparison with the results we obtained for UMa II and Seg 1 is less clear-cut, as we found that the exact values of the Chabrier IMF were rejected, although neighboring grid points were not rejected.

The results that we found above for a lognormal IMF in Ret II and Tri II are also fully consistent with the subset of galaxies presented in Gennaro et al. (2018a) that have (system) IMF parameter values that have the Chabrier values within the 68% CIs of their posteriors. The best-fit values we found for lognormal IMFs for Ret II and UMa II are similar to those of Boo I that we found in Paper I (namely, $M_c = 0.17$ $\mathcal{M}_\odot$, $\sigma = 0.49$), with Ret II having the largest value for $\sigma$ of the three ($\sigma_{\text{RetII}} = 0.65$).

## 6. Conclusions

We analyzed deep HST/ACS imaging in two filters, F814W and F606W, reaching down to apparent magnitudes that are equivalent to $\lesssim 0.2\ \mathcal{M}_\odot$, for four UFD galaxies—Ret II, UMa II, Tri II, and Seg 1—with the aim of constraining the low-mass stellar IMF ($M < 1\ \mathcal{M}_\odot$). The resulting catalogs of likely member sources for two of these galaxies—Ret II and UMa II—are large enough that we could carry out ABC-MCMC analyses of the color–magnitude data and determine "best-fit" values for the parameters of three forms of the IMF, namely, a single power law, a broken power law, and a lognormal function. All four galaxies have a sufficiently large catalog that we could model synthetic stellar populations, in the form of apparent-magnitude distributions in each of two filters, with assumed IMF parameter values and binary fractions on a grid, and carry out KS tests between the observed and synthetic data. These KS tests allowed us to identify which combinations of parameter values could be rejected, as the observed and synthetic data were not consistent with having been drawn from the same distribution. We found the following:

1. The results from the ABC-MCMC analyses, given in Table 3, show that, for the lognormal form of the IMF, the best-fit values for both Ret II and UMa II are close to the parameter values for the Milky Way IMF, which in turn are well within the 68% CI for each system. The results for the broken power-law form, however, are further from the Milky Way values, and indeed, the best-fit value for the slope below the break mass in UMa II is particularly dissimilar to that for the Milky Way, favoring a distinctly more bottom-heavy IMF. The Milky Way values for the slope below the break mass lie just outside the 68% CI for Ret II (and, indeed, are not in tension if the uncertainties on the Milky Way values are considered), and the slope above the break mass is entirely consistent with the Milky Way values. The results for the single power-law form are not straightforward to interpret, as the low-mass stellar populations of the Milky Way are not well described with a single power law.[15]

2. The KS tests show that several different combinations of functional forms of the IMF and parameter values can be rejected. There are also a number of combinations that cannot be rejected in our conservative limit (i.e., $p > 0.05$ for both photometric filters for at least one of the 1000 synthetic data realizations of a given combination of IMF parameters and binary fraction). The results of tests using synthetic populations with IMF parameter values close to those of the Milky Way are summarized in Table 4, and show that an underlying IMF close to that of the Milky Way cannot be rejected, even if, in the case of UMa II, parameter values exactly matching those reported for the Milky Way are rejected. There does not appear to be any systematic dependence of the IMF on properties of the galaxy. Indeed, there is significant overlap in the regions of parameter space that are not rejected for all the galaxies, indicating that it is plausible that these four galaxies all have the same IMF.

3. The ABC-MCMC analysis suggests that the low-mass IMF in UMa II is relatively bottom heavy compared with that of the stellar populations of the Milky Way. This comes with the caveat that there are two $z \sim 0.5$ galaxy clusters located within a few arcminutes of the center of UMa II, which could contribute additional background contamination, compared to the fields of the other UFD galaxies in our sample. An increasing level of contamination from these clusters as a function of apparent magnitude (following the galaxy luminosity function) could cause the IMF in UMa II to appear more bottom heavy.

It is clear that larger samples than in the present analysis, with similarly deep photometry, would aid the determination of the low-mass IMF. This should be possible with the next generation of space telescopes, and indeed, there are already a number of JWST programs that aim to constrain the low-mass IMF of UFD galaxies (e.g., Weisz et al. 2023), and there will likely be similar programs with the Nancy Roman Space Telescope. Increasing the areal coverage comes with the cost of an increased possibility of contamination, given the surface density profiles of these faint targets. Sufficiently deep and precise proper motions from repeated imaging could mitigate this, but would require multiyear baselines. Ultimately, the low luminosities of UFD galaxies set fundamental limits on the sample size one could hope to obtain.

The use of three or more photometric passbands would also be beneficial to future analyses, as this would allow the creation of color–color diagrams to aid the identification and removal of background galaxies. As noted above, we suspect that our reliance on PSF quality-of-fit statistics and color–magnitude cuts led to a higher level of contamination from faint background galaxies in the UMa II fields than for the other UFD galaxies, which could explain why our analysis of the data favors a bottom-heavy IMF for UMa II. Incorporation of an additional passband could help to discern whether the low-mass IMF of UMa II truly does differ from that of the low-mass stellar populations of the Milky Way.

### Acknowledgments

C.F. and R.F.G.W. are grateful for support through the generosity of Eric and Wendy Schmidt, by recommendation of the Schmidt Futures program. C.F. acknowledges support by the NASA FINESST grant (80NSSC21K2042). R.F.G.W. thanks her sister, Katherine Barber, for her support. We thank the referee for the insightful comments that helped improve the paper.

The HST data were observed as part of Treasury Program GO-14734 (PI Kallivayalil). Support for this program was provided by NASA through grants from the Space Telescope Science Institute, which is operated by the Association of Universities for Research in Astronomy, Incorporated, under NASA contract NAS5-26555. The catalogs for Ret II, UMa II, Tri II, and Seg 1 will be available at doi:10.17909/b5gn-6e22, and the Boo I catalog (see Appendix A) is available at doi:10.17909/g37r-1r51. The imaging data of Ret II, UMa II, Tri II, and Seg 1 are available via doi:10.17909/mwdx-da66. This research has made use of the SIMBAD database, operated at Strasbourg astronomical Data Center (CDS), Strasbourg, France, together with data from the European Space Agency (ESA) mission Gaia (https://www.cosmos.esa.int/gaia), processed by the Gaia Data Processing and Analysis Consortium (DPAC, https://www.cosmos.esa.int/web/gaia/dpac/consortium). This research also made use of the cross-match service provided by CDS, Strasbourg. Funding for the DPAC has been provided by national institutions, in particular the institutions participating in the Gaia

---

[15] Note that the data set for low-mass stars in the Milky Way that is modeled with a single power law by Geha et al. (2013) is, in the original analysis of Bochanski et al. (2010), best fit by a lognormal IMF.

Multilateral Agreement. We also utilized the NASA IPAC IRSA Galactic Dust and Extinction web query at https://irsa.ipac.caltech.edu/applications/DUST/ and the SVO filter profile service at http://svo2.cab.inta-csic.es/theory/fps/. This research made use of Photutils, an Astropy package for detection and photometry of astronomical sources (Bradley et al. 2023).

*Facility*: HST.

*Software*: DAOPHOT-II (Stetson 1992), astropy (Astropy Collaboration et al. 2013, 2018, 2022), photutils (Bradley et al. 2023), DrizzlePac (STSCI Development Team 2012), emcee (Foreman-Mackey et al. 2013), Isochrones (Morton 2015), matplotlib (Hunter 2007), NumPy (Harris et al. 2020), SciPy (Virtanen et al. 2020), pandas (McKinney 2010), pygtc (Bocquet & Carter 2016), TensorFlow (Abadi et al. 2015).

## Appendix A
## Proper-motion Data for Ursa Major II

As noted in the main text, the two most enriched stars in the catalog of likely members analyzed by Kirby et al. (2013) are outliers in proper motion compared to the remaining sample. This is illustrated in Figure A1, based on Gaia EDR3 proper motions (Gaia Collaboration et al. 2016, 2021). We cross-matched the catalogs using the CDS X-Match service, with a 1″ matching radius. The two most metal-rich stars (shown by points with the lightest colors) are offset in proper motion from both the other stars plotted and the systemic proper motion of UMa II ($\mu_{\alpha*} = 1.73^{+0.02}_{-0.02}$ mas yr$^{-1}$, $\mu_\delta = -1.90^{+0.02}_{-0.02}$ mas yr$^{-1}$) from Battaglia et al. (2022). We therefore truncated the metallicity distribution at [Fe/H] = −1.5, between the second-most and thirdmost metal-poor stars in the sample.

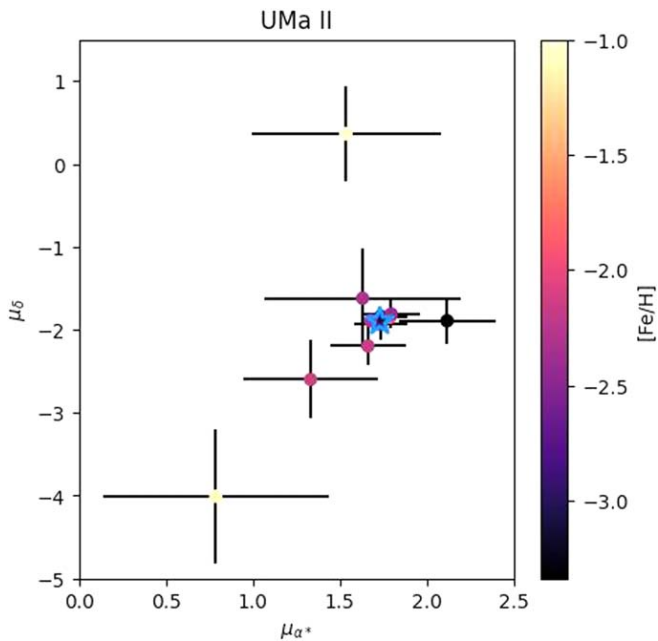

**Figure A1.** Proper motions and associated errors from Gaia EDR3 (Gaia Collaboration et al. 2016, 2021) of likely UMa II member stars, with membership taken from Kirby et al. (2013). The systemic proper motion of UMa II from Battaglia et al. (2022) is indicated by the blue star, while the points for the individual stars are color-coded by the metallicity estimate in Kirby et al. (2013). It can be seen that the two stars with the highest metallicities (∼ −1 dex, shown with the lightest colors) are outliers in proper motion.

## Appendix B
## Verification of Our Implementation of the KS Tests

The approach that we used to undertake and interpret the KS tests was described in Section 3.4. We verified that the results obtained following this procedure were valid by applying it to synthetic populations of assumed IMF parameters and binary fractions. We generated "observed" samples of ∼550 sources, adopting physical parameters (such as distance and metallicity) that matched those of Tri II. We first adopted an IMF with one of the three functional forms used in the main analysis and investigated which values of the parameters were rejected by KS tests with a parent population of matching functional form of the IMF, and with parameter values from the grid described in the main text (Section 3.4). We then investigated the outcomes of KS tests where the "observed" population and the simulated populations have a different form of the IMF, as described in Section B.2 below. We also applied KS tests to the data for Boo I from Paper I (Section B.3) and checked that we obtained consistent results to those of Paper I, which were obtained using the same ABC-MCMC algorithm as in the present paper.

### B.1. Matched Functional Forms of the IMF

The results of our tests with each of the three forms of the IMF, namely, single power-law, broken power-law, and lognormal functions, are described below.

#### B.1.1. Single Power Law

The "observed" population was generated with a binary fraction of 0.6 and a power-law IMF with slope equal to the Salpeter value ($\alpha = -2.35$). We adopted the same grid of slope values and binary fractions for the synthetic test populations as was given in Section 3.4. The results, together with the grid values, are presented in the leftmost panel of Figure B1. It can be seen that, for each of the three binary fractions tested, the Salpeter value for the slope (which is not a grid point) is bracketed by grid points that are not rejected, indicating that this assumed value for the "observed" population is indeed not rejected by the KS tests.

#### B.1.2. Broken Power Law

We generated an "observed" population with binary fraction $f_f = 0.3$ and a broken power-law IMF with parameter values equal to those reported by Kroupa (i.e., $\alpha_{1K} = -1.3$, $\alpha_{2K} = -2.3$). Again, we adopted the grid of parameter values used for the real data, given in Section 3.4, so again, the grid does not include the exact Kroupa values, but does include the very similar values of $\alpha_1 = -1.25$ and $\alpha_2 = -2.25$. The results, together with the grid values, are presented in the middle panel of Figure B1, where it can be seen that this closest grid point to the Kroupa values is not rejected, independent of the binary fraction.

It is interesting to note that, while most combinations of parameter values that were not rejected have $|\alpha_2| > |\alpha_1|$, i.e., a steeper slope for masses above the break mass than for those below it, some combinations that are not rejected have $\alpha_1 = \alpha_2$, or even $|\alpha_2| < |\alpha_1|$. This range of behaviors that is not rejected demonstrates that when the population is sampled by only a small data set, as is the case for most of the sample of UFD galaxies studied here, the existence of nonrejected slopes in a

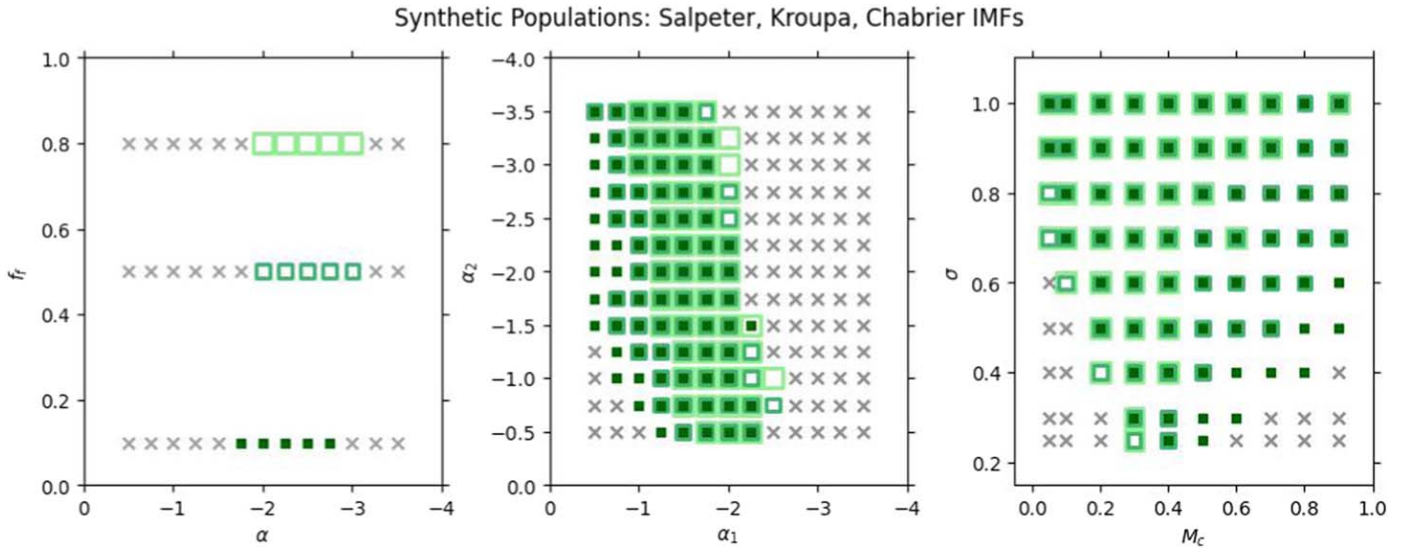

**Figure B1.** The results of the KS tests for grids of the different functional form of the IMF: single power law (left), broken power law (middle), and lognormal (right). The symbols follow the scheme given in the caption to Figure 7, which we recap here for convenience. Each point on the plots indicates the values of the IMF parameters and binary fraction for one of the synthetic populations compared with the "observed" population. We investigated three binary fractions, $f_f = 0.1$, 0.5, and 0.8, indicated by small filled green squares, open medium-size green squares, and large open green squares, respectively. Labeling grid points with these symbols indicates that the combination of that binary fraction and the IMF parameter values at that grid point are not rejected (i.e., the $p$-value from the KS test is above 0.05 for the photometry in both the F606W and F814W filters, for at least one of the 1000 realizations of synthetic stellar populations that were generated with that combination of binary fraction and IMF parameter values). The gray crosses in the middle and left panels indicate grid points where every realization of populations with those IMF parameter values are rejected for each of the three binary fractions, while in the left panel, where the vertical axis corresponds to the binary fraction, the gray crosses indicate that the plotted value is rejected.

region of parameter space far from the Kroupa values (such as where $|\alpha_2| < |\alpha_1|$), in addition to the Kroupa values, cannot be seen as evidence for tension between the underlying IMF of the stellar population and the Kroupa IMF.

### B.1.3. Lognormal

Finally, we generated an "observed" population drawn from a Chabrier IMF ($M_{cc} = 0.2\ \mathcal{M}_\odot$, $\sigma_c = 0.55$) and a binary fraction equal to 0.4. The grid of parameter values for the KS tests again follows that given in Section 3.4, and the results are shown in the rightmost panel of Figure B1. Once again, a range of values for each of $\sigma$, $M_c$, and $f_f$ cannot be rejected, including those closest to the input values of the "observed" population ($M_c = 0.2\ \mathcal{M}_\odot$ with $\sigma = 0.5$ and $f_f = 0.1$ or 0.5, or with $\sigma = 0.6$ with $f_f = 0.1$ or 0.5; note that $\sigma = 0.55$ is not included in the grid). The region of parameter space that is not rejected here is significantly larger than for the KS tests on the real, observed data, and extends to combinations of $M_c$ and $\sigma$ that are far from the Chabrier values. We briefly comment below on the possible role of contamination in setting the extent of the regions of parameter spaces that are not rejected.

### B.1.4. Size of Parameter Space that Is Not Rejected

The extent of parameter space that is not rejected in these tests is larger than that for the case of the KS tests performed with the real observed data. The region of parameter space that is not rejected for the synthetic "observed" population is shifted slightly to the left in the plot showing $\alpha_1$ versus $\alpha_2$ (i.e., toward smaller $|\alpha_1|$) when compared to the results of the KS tests on the observed data. Similarly, the majority of the combinations with $M_c > 0.4\ \mathcal{M}_\odot$ that are not rejected in Figure B1 are shown as rejected in the corresponding figures for the observed data.

The tests presented in this section use entirely synthetic populations and so lack contamination by nonmember sources such as stars in the Milky Way or background, barely resolved galaxies.

It may be expected that this additional means of deviation between the actual data and the synthetic data would increase the extent of the parameter space that does not provide a good match and is hence rejected. A low level of contamination should have a similar effect on the KS tests, through distortion of the observed apparent magnitude distribution, as does the effect of small sample size, or the uncertainties in distance and extinction that are incorporated in the different synthetic populations. Thus, we would expect the true underlying IMF not to be rejected. However, a large contribution from contaminating sources could lead to the rejection of the true IMF.

We tested this proposal by successively replacing increasingly larger portions of a synthetic Tri II-like stellar population with "contamination." We first generated the Tri II-like population (as in the earlier tests) using a Kroupa IMF and a binary fraction of 0.3. We generated the contamination by drawing F814W magnitudes between the limits F814W = 23.5 to F814W = 26.88 from a distribution where the probability linearly increased toward fainter magnitudes, such that F814W = 26.88 was 500 times more probable than F814W = 23.5. We generated a color for each magnitude from a uniform distribution between the minimum and maximum colors of the "stars" in the synthetic population that were fainter than F814W = 24.0, and we required that this color was within the bounds set by the fiducial isochrone for Tri II. The corresponding F606W magnitude is then the sum of this color and the F814W magnitude, and again is required to be within the apparent magnitude bounds. The level of contamination was set at 5% initially, then increased in increments of 5% until the Kroupa IMF was rejected. As the contamination level increased, we found that first the combination of a Kroupa IMF with $f_f = 0.8$ was rejected, then the combination with $f_f = 0.5$ was rejected, and then, once the contamination reached 35%, the Kroupa IMF was rejected for all three binary fractions.

We repeated these tests of the effects of contamination using a Chabrier IMF and a 30% binary fraction to generate the Tri II-like synthetic stellar population and found consistent results.

### B.2. Cross Tests with Unmatched Functional Forms of the IMF

In addition to the tests described above, we performed a test where the functional form of the IMF adopted for the KS test grid differed from that of the IMF used to generate the "observed" population. This allowed us to investigate the effect of a mismatch between the underlying mass distribution (i.e., the ground truth) and the functional form being tested. We adopted the same "observed" stellar population that we used for the matched broken power-law and lognormal IMF tests presented in Section B.1.

#### B.2.1. Broken Power-law "Observed" Population, Lognormal Grid

The results of comparing an "observed" population created assuming a broken power-law form of the IMF, with parameter values chosen to match the Kroupa IMF, to a grid of synthetic populations generated using a lognormal form of the IMF are shown in Figure B2. It can be seen that the extent of the parameter space that is not rejected is reduced compared to that when both the "observed" population and the test grid were drawn from a lognormal IMF. Notably, only combinations with binary fractions $f_f = 0.1$ are not rejected for characteristic masses $M_c \geqslant 0.4\ \mathcal{M}_\odot$, and there are fewer combinations with $M_c \geqslant 0.4\ \mathcal{M}_\odot$ that are not rejected, compared with the case of matched IMFs (e.g., the middle panel of Figure B1). In contrast, more combinations of parameters for $M_c \leqslant 0.2\ \mathcal{M}_\odot$ are not rejected, including combinations with $M_c = 0.1\ \mathcal{M}_\odot$, and $\sigma = 0.5$. It is evident that the grid point closest to the Chabrier parameters would not be rejected, which is to be anticipated given that the Chabrier and Kroupa IMFs are sufficiently similar that they are both used to describe the same low-mass stellar populations of the Milky Way.

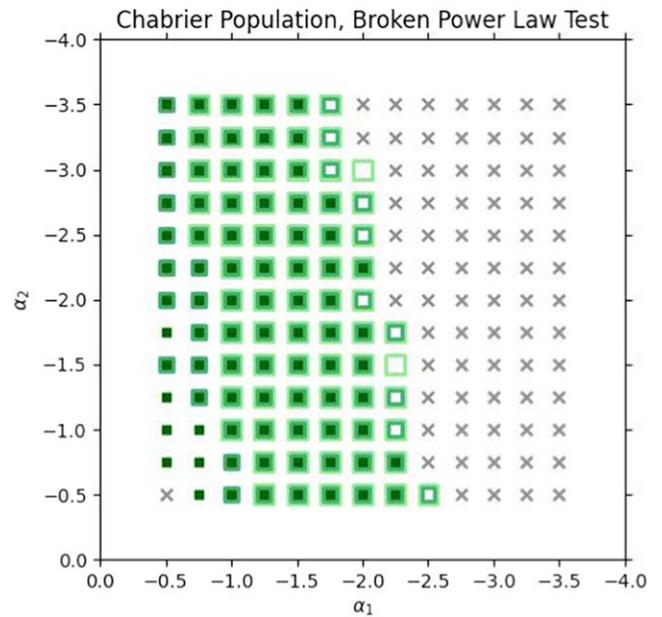

**Figure B3.** The results of the KS tests from the broken power-law parameter grid applied to a "method" population generated with a Chabrier lognormal IMF. The color-coding and symbols follow the scheme described in the captions of Figure B1 and Figure 7.

#### B.2.2. Lognormal "Observed" Population, Broken Power-law Grid

Figure B3 shows the regions of parameter space for a broken power-law IMF that are not rejected when compared to an "observed" population generated with a lognormal IMF with parameter values equal to those of the Chabrier IMF.

Sections of the nonrejected parameter space for $\alpha_1$ and $\alpha_2$ are quite similar to that of the broken power-law "observed" population (compare with the right-hand panel of Figure B1 above), albeit with generally different binary fractions. Here, there are no combinations of parameters with $|\alpha_1| > 2.50$ that are not rejected, and there are more combinations with $|\alpha_1| \leqslant 1$ that are not rejected. The range of parameters that are not rejected has shifted toward the lower left-hand corner of the plot, compared to the case of matched IMFs. However, it would still be concluded that an IMF similar to that of Kroupa (which describes the same Milky Way population as the Chabrier IMF used to generate the "observed" population) would not be rejected, as expected.

### B.3. Boo I

We further tested (and verified) that the results of KS tests carried out following our procedure are consistent with our earlier findings for Boo I, carried out using an ABC-MCMC approach and reported in Paper I. The results of the KS tests are shown in Figure B4, following the format of Figure 7. As may be seen in the left-hand panel, single power-law slopes of $\alpha = -1.75, -2.00$, or $-2.25$ cannot be rejected. This is entirely consistent with the best-fit single power-law slope found in Paper I ($\alpha = -1.95$). Continuing to the case of the broken power-law IMF (middle panel), the best-fit slopes found in Paper I were $\alpha_1 = -1.67$ and $\alpha_2 = -2.57$, and the grid points nearest these values cannot be rejected. Finally, the grid points closest to the best-fit parameter values found in Paper I for the case of the lognormal IMF ($M_c = 0.17 \mathcal{M}_\odot$, and $\sigma = 0.49$) also cannot be rejected (see the right-hand panel). Further, the grid points close to each of the Kroupa and Chabrier IMF parameter values are not rejected, in agreement with our earlier conclusions that a Milky Way IMF was compatible with the data for Boo I.

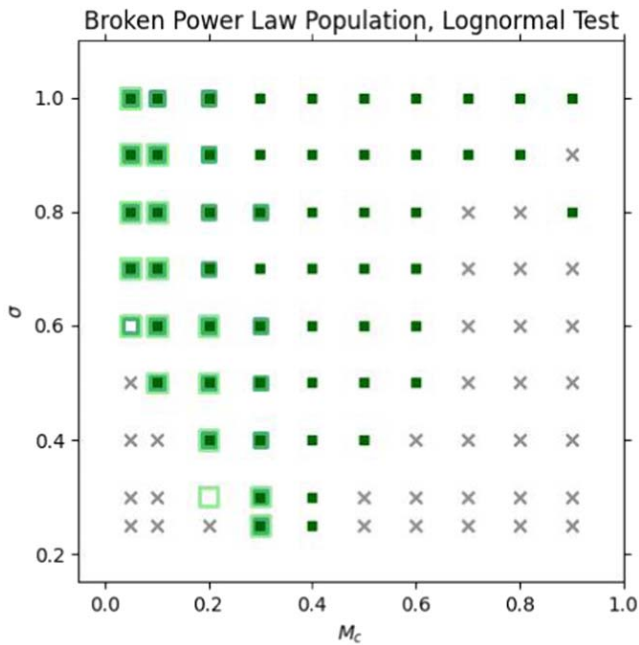

**Figure B2.** The results of the KS tests from the lognormal parameter grid applied to an "observed" population generated with a Kroupa broken power-law IMF. The color-coding and symbols follow the scheme described in the captions of Figure B1 and Figure 7.

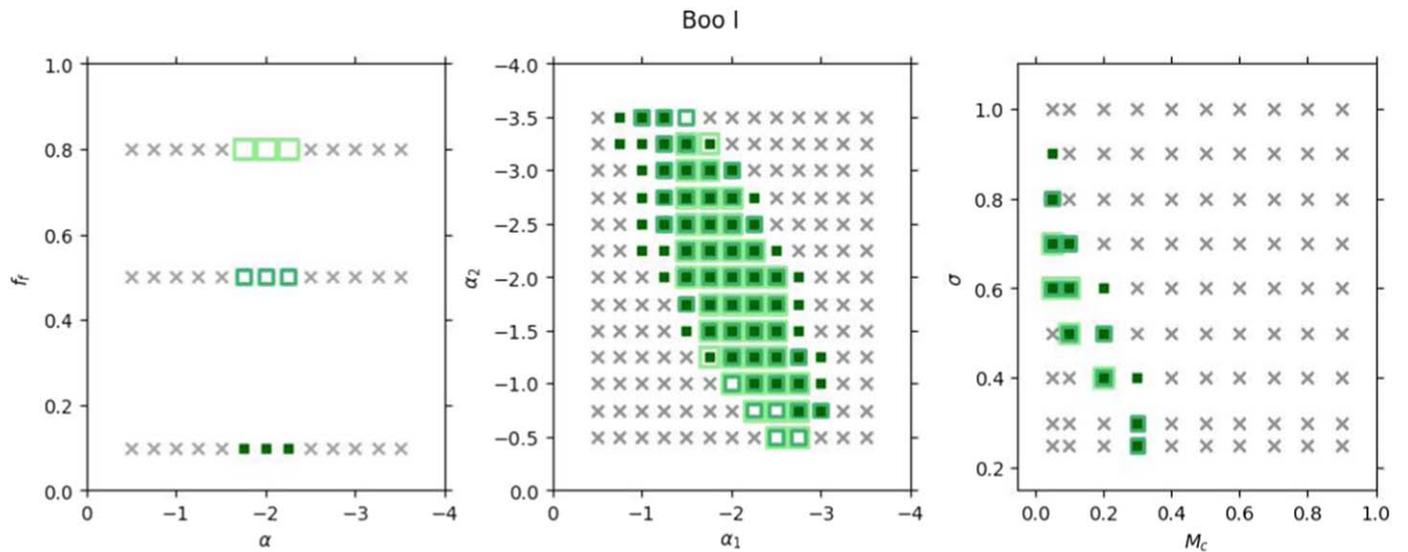

**Figure B4.** The results of the KS tests for the single power-law (left), broken power-law (middle), and lognormal (right) forms of the IMF for Boo I. The color-coding and symbols follow the scheme described in the captions of Figure B1 and Figure 7.

## ORCID iDs

Carrie Filion ⓘ https://orcid.org/0000-0001-5522-5029
Rosemary F. G. Wyse ⓘ https://orcid.org/0000-0002-4013-1799
Hannah Richstein ⓘ https://orcid.org/0000-0002-3188-2718
Nitya Kallivayalil ⓘ https://orcid.org/0000-0002-3204-1742
Roeland P. van der Marel ⓘ https://orcid.org/0000-0001-7827-7825
Elena Sacchi ⓘ https://orcid.org/0000-0001-5618-0109